\documentclass[]{aa}
\usepackage{graphicx}
\usepackage{txfonts}
\usepackage{longtable}
\usepackage{xcolor}
\usepackage{pdfpages}
\usepackage{verbatim}

\bibpunct{(}{)}{;}{a}{}{,}

\begin{document} 

\newcommand{\hi}{\mbox{H\,{\sc i}}}
\newcommand{\zabs}{$z_{\rm abs}$}
\newcommand{\zmin}{$z_{\rm min}$}
\newcommand{\zmax}{$z_{\rm max}$}
\newcommand{\zq}{$z_{\rm q}$}
\newcommand{\zg}{$z_{\rm g}$}
\newcommand{\kms}{km\,s$^{-1}$}
\newcommand{\cmsq}{cm$^{-2}$}
\newcommand{\degree}{\ensuremath{^\circ}}
\newcommand{\Msun}{$M_{\odot}$} 
\newcommand{\mgii}{\mbox{Mg\,{\sc ii}}} 
\newcommand{\mgiia}{\mbox{Mg\,{\sc ii}$\lambda$2796}}
\newcommand{\mgiib}{\mbox{Mg\,{\sc ii}$\lambda$2803}}
\newcommand{\mgiiab}{\mbox{Mg\,{\sc ii}$\lambda\lambda$2796,2803}}
\newcommand{\lapp}{\mbox{\raisebox{-0.3em}{$\stackrel{\textstyle <}{\sim}$}}}
\newcommand{\gapp}{\mbox{\raisebox{-0.3em}{$\stackrel{\textstyle >}{\sim}$}}}

\newcommand{\lux}{Observatoire de Paris, LUX, Coll\`ege de France, CNRS, PSL University, Sorbonne University, 75014, Paris, France --- \email{francoise.combes@obspm.fr} \label{lux}}
\newcommand{\iac}{Instituto de Astrofísica de Canarias, Vía Láctea, S/N, E-38205, La Laguna, Spain \label{iac}}
\newcommand{\dau}{Departamento de Astrofísica, Universidad de La Laguna, E-38206, La Laguna, Spain \label{dau}}
\newcommand{\madrid}{Observatorio Astronómico Nacional (OAN-IGN)-Observatorio de Madrid, Alfonso XII, 3, 28014, Madrid, Spain \label{madrid}}
\newcommand{\firenze}{INAF - Osservatorio Astrofisico di Arcetri, Largo Enrico Fermi 5, I-50125 Firenze, Italy \label{firenze}}
\newcommand{\onsala}{Department of Space, Earth and Environment, Chalmers University of Technology, Onsala Space Observatory, 439 92, Onsala, Sweden \label{onsala}}
\newcommand{\bologna}{INAF - Istituto di Radioastronomia, Via P. Gobetti 101, 40129, Bologna, Italy \label{bologna}}
\newcommand{\athens}{Section of Astrophysics, Astronomy, and Mechanics, Department of Physics, National and Kapodistrian University of Athens, Panepistimioupolis Zografou, 15784 Athens, Greece \label{athens}}
\newcommand{\koln}{Physikalisches Institut, Universität zu Köln, Zülpicher Str. 77, 50937, Cologne, Germany \label{koln}}
\newcommand{\kolnb}{Max-Plank-Institut für Radioastronomie, Max-Planck-Gesellschaft, Auf dem Hügel 69, 53121, Bonn, Germany \label{kolnb}}
\newcommand{\iram}{Institut de Radioastronomie Millimétrique (IRAM), 300 Rue de la Piscine, 38400 Saint-Martin-d'Hères, France \label{iram}}
\newcommand{\eso}{European Southern Observatory, Alonso de Córdova, 3107, Vitacura, Santiago, 763-0355, Chile \label{eso}}
\newcommand{\esob}{Joint ALMA Observatory, Alonso de Córdova, 3107, Vitacura, Santiago, 763-0355, Chile \label{esob}}
\newcommand{\taipei}{Institute of Astronomy and Astrophysics, Academia Sinica, 11F of AS/NTU Astronomy-Mathematics Building, No.1, Sec. 4, Roosevelt Rd, Taipei, 106319, Taiwan \label{taipei}}
\newcommand{\leiden}{Leiden Observatory, Leiden Univ., PO Box 9513, 2300 RA, Leiden, The Netherlands \label{leiden}}
\newcommand{\bonn}{Transdisciplinary Research Area (TRA) ‘Matter’/Argelander-Institut f\"ur Astronomie, University of Bonn \label{bonn}}
\definecolor{green}{rgb}{0,0.4,0}

\titlerunning{Molecular tori at high resolution}
\authorrunning{F. Combes, A. Audibert, S. Garcia-Burillo et al. }

   \title{High resolution mapping of molecular tori with ALMA }
   \author{
        F. Combes\inst{\ref{lux}}
        \and A. Audibert\inst{\ref{iac},\ref{dau}}
        \and S. Garcia-Burillo\inst{\ref{madrid}}
        \and L. Hunt\inst{\ref{firenze}}
        \and S. Aalto\inst{\ref{onsala}}
        \and V. Casasola\inst{\ref{bologna}}
        \and K. Dasyra\inst{\ref{athens}}
        \and A. Eckart\inst{\ref{koln},\ref{kolnb}}
        \and M. Krips\inst{\ref{iram}}
        \and S. Martin\inst{\ref{eso},\ref{esob}}
        \and S. Muller\inst{\ref{onsala}}
        \and K. Sakamoto\inst{\ref{taipei}}
        \and P. van der Werf\inst{\ref{leiden}}
        \and S. Viti\inst{\ref{leiden},\ref{bonn}}
           }

    \institute{\lux \and \iac \and \dau \and \madrid \and \firenze \and \onsala  \and \bologna \and \athens \and \koln \and \kolnb \and \iram \and \eso \and \esob \and \taipei \and \leiden \and \bonn
   }

   \date{Received: September 2025; accepted: November 2025}
 
  \abstract {
 Recent high resolution mapping of the circum-nuclear regions of Active Galactic Nuclei (AGN) has revealed the
 existence of geometrically thin nuclear disks, in general randomly oriented with 
 respect to their galaxy hosts.
 These molecular tori have typical radii of 10~pc, and contain a 
 few 10$^7$ M$_\odot$ of H$_2$, with H$_2$ column
 densities between 10$^{23}$ and 10$^{25}$ cm$^{-2}$. We mapped two of the most massive of these molecular
 tori with higher resolution, in order to unveil their morphology and kinematics, their possible warp and
 clumpiness, and derive their stability and life-time. We used the highest resolution possible with
 ALMA (16~km baseline) in Band 7, taking into account for mapping CO(3-2) and HCO$^+$(4-3)
 the compromise between sensitivity and resolution. 
 New features are discovered at the high resolution, obtained with a beam of 0.015\arcsec, equivalent to
$\sim$ 1~pc scale, at their $\sim$ 15~Mpc distance. The molecular torus in NGC~613 appears like a ring, depleted in molecular gas near the center. The depletion region is displaced by 3~pc toward the NW from the AGN position, meaning some $m=1$ asymmetry in the torus.
 The molecular torus in NGC~1672 has a different position angle  from previous
 lower-resolution observations,
 and is edge-on, revealing a geometrically very thin torus  (axis ratio 6.5 to 10), with a clear warp. 
 This confirms that the classical model of a simple geometrically thick dusty torus is
 challenged by high resolution observations. The nuclear disks
 appear clumpy, and slightly lopsided. The molecular outflow in NGC~613 is now resolved out.
 Well inside the sphere of influence of the black holes (BH), we are now able to determine more accurately their
mass, for those Seyfert spiral galaxies, in a region of the M-sigma relation where the scatter is maximum.
}
   

\maketitle
%
\section{Introduction}
\label{sec:intro}

How are active galaxy nuclei (AGN) fueled, and how do they regulate star formation? In recent years, observations at 100~pc scales of the molecular gas have brought great progress in the question of how AGN are fueled dynamically in galaxies \citep[e.g.][]{Garcia-Burillo2005, Combes2013, Combes2014, Garcia-Burillo2021, Garcia-Burillo2024, Goesaert2025}, and how the energy generated by the AGN can, in turn, regulate its gas accretion, through molecular outflows \citep[e.g.][]{Cicone2014, Garcia-Burillo2014, Fiore2017, Fluetsch2019, Lutz2020,Gorski2024,Esposito2024}. 
This star formation regulation between feeding and feedback is well illustrated in numerical simulations of embedded bars, spirals, and m=1 instabilities \citep[e.g.][]{Hopkins2010, Combes2012, Hopkins2024}, and also corresponding feedback 
\citep{Dubois2013, Gabor2014, Ward2022, Koudmani2022}.
ALMA has recently mapped in molecular lines and dust continuum some molecular tori, a key element in the AGN unification paradigm \citep[e.g.][]{Garcia-Burillo2016, Combes2019, Impellizzeri2019,
Garcia-Burillo2019,Kameno2020,Imanishi2020,Uzuo2021,Alonso-Herrero2023}. After having mapped the nuclear regions of several nearby Seyferts with a few 10's parsec resolution, our view of molecular tori is more complex than the iconic model brought by optical, X-ray and radio observations \citep[e.g.][]{Antonucci1985, Urry1995}.  The torus might be unstable, warped, and decoupled from larger scales. The 10~pc scale is a pivotal region of transition between the gas being driven to the center dynamically to feed the nucleus, and the AGN feedback on the host galaxy, via molecular outflows.

These 10~pc-scales are the frontiers separating the BLR (Broad Line Regions) from the NLR (Narrow Line Regions), the main components observed for decades in the optical AGN spectra. The torus is supposed to obscure the BLR in type 2 AGN, and is believed to collimate the AGN radiation, which produces a bi-conical structure \citep[e.g.][]{Malkan1998}. Therefore, dust  in this NLR region is observed with a polar structure \citep{Gratadour2015, Asmus2016, Leftley2019, Asmus2019, Haidar2024}. 
 Multi-wavelength data have converged to provide clues on the circum-nuclear obscuring structure
 \citep[e.g.][]{Ramos2017, Hoenig2019}.
The expected torus is so small (1-30~pc in size) that it was impossible to see it until recently, where we have detected for the first time CO(6-5) emission in a 10~pc-diameter torus in the Seyfert 2 NGC 1068 with ALMA \citep{Garcia-Burillo2016}. \citet{Imanishi2016} found there strong HCO$^+$ and HCN emissions, tracing dense gas.

In the present work, we report high spatial resolution (1~pc) mapping of the molecular gas through the CO(3-2) line, in two spiral galaxies, where dense molecular tori were found \citep{Combes2019}, in order to analyze further their morphology and kinematics.
The paper is structured as follows: our sample is briefly described in Sect. \ref{sec:samp}. Section \ref{sec:obs} presents the details of ALMA observations. The morphology and kinematics results are quantified in Section \ref{sec:res}. We discuss in
Sec. \ref{sec:disc} the inferences on feeding and feedback processes for these low-luminosity AGN. Section \ref{conclu} summarizes our conclusions.
Throughout the present paper, the velocity scale is defined with respect to redshifts indicated in Table
\ref{tab:samp}, if not mentioned otherwise. With the distances adopted in Table \ref{tab:samp}, 1\arcsec\ = 83~pc and 55~pc for
NGC~613 and NGC~1672 respectively.

\section{The sample}
\label{sec:samp}
The galaxies studied are two of the most gas-concentrated objects in the NUGA (NUclei of GAlaxies) sample of nearby low-luminosity AGN that are for a large fraction barred galaxies \citep[e.g.][]{Garcia-Burillo2003, Casasola2011}. 
They were observed with ALMA covering different tracers (CO(3-2), CO(6-5), CI and dense gas tracers, like
HCN(4-3), HCO$^+$(4-3), CS(7-6)), and a large range of spatial resolutions. Our main goal is to study the dynamics and structures in the central kpc down to the sphere of influence of the black hole (with the highest reolution of 1~pc), tracing the kinematics in the galaxy disks, possible bars and nuclear rings, computing gravity torques to understand the fueling mechanisms, and tracking feedback outflows \citep{Combes2013, Combes2014, Combes2019, Audibert2019, Audibert2021}.

  NGC~613 is a typical barred spiral with an inner Lindblad resonance (ILR) ring 
  of radius 3.5\arcsec\ or 300~pc, and inside the ring, 
  a circumnuclear disk (CND) of radius 1\arcsec\ = 83~pc. \citet{Miyamoto2017} found a radio continuum jet at 95~GHz
  (PA=20$^\circ$) corresponding to the radio-cm jet \citep{Hummel1992}. \citet{Audibert2019}
have detected a nuclear molecular outflow (size $\sim$ 25~pc) along the jet,
  and the molecular torus is located inside a nuclear trailing spiral, fueling the nucleus. The main disk is
  rather face-on (i = 36$^\circ$, PA = 122$^\circ$ from HyperLeda, see reference link in the acknowledgements), while the torus is kinematically decoupled and
   conspicuous in CO(3-2) and HCO$^+$(4-3) \citep{Audibert2019}. Water masers are present in the nucleus \citep{Kondratko2006}. 
  
  NGC~1672 is a strongly barred Sy2, with a high global star formation activity, SFR= 2.7~M$_\odot$/yr
  \citep{Kewley2000}. The AGN is hidden in a
  Compton-thick nucleus \citep{deNaray2000}. Chandra detected a hard X-ray emission in the nucleus \citep{Jenkins2011}. 
  Inside the ILR nuclear ring of radius 5\arcsec\ =275~pc, some thin filaments join to the torus, kinematically
  decoupled from the galaxy,
  with a large velocity gradient inside 100~pc. While the main galaxy disk is almost face-on 
  (i = 29$^\circ$, PA = 155$^\circ$ from HyperLeda), the torus appears much more inclined 
   (i = 66$^\circ$, PA = 90$^\circ$) \citep{Combes2019}.
  We summarize the galaxy properties in Table \ref{tab:samp}.

\begin{table*}[!htb]
      \caption[]{Targets' physical parameters}
         \label{tab:samp}
            \begin{tabular}{l c c c c c c c c c c }
            \hline
            \noalign{\smallskip}
   Name&   Type &    D   &SFR    & log(L$_X$)   & log(L$_{1.4GHz}$)& S(CO)$_{32}$ & Bar  & z &RA & DEC \\
            \noalign{\smallskip}
       &        &  Mpc & M$_\odot$/yr& erg/s & W/Hz & Jy \kms & PA($^\circ$)      &  & ICRS & ICRS \\
            \hline
            \noalign{\smallskip}
N613 & Sy- SB(rs)bc  & 17.2& 5.3  & 41.2  & 21.8   &  56. & 127  & 0.00494 & 01:34:18.189 &  -29:25:06.59\\
N1672& Sy 2- SB(s)b  & 11.4& 3.1  & 38.4  & 19.9   & 80.  &  97  & 0.00444 & 04:45:42.496 &  -59:14:49.91\\
            \noalign{\smallskip}
            \hline
            \end{tabular}
            \begin{list}{}{}
\item -- D are the median values of z-independent distances from NED \citep{Steer2017}\\
  -- SFR are derived from infrared luminosities (NED)\\
 -- L$_X$ is from 2-10~keV INTEGRAL, Rosat and/or Chandra archives\\
 -- The CO(3-2) integrated fluxes are from the molecular torus only \citep{Combes2019}\\
  -- PA of bars are from \cite{Jungwiert1997} for NGC~613, and from \cite{Jenkins2011} for
  NGC~1672\\
  -- The RA-DEC positions are the adopted centers for each galaxy, derived from the continuum point sources
  detected with ALMA, with an error bar of $\sim$ 0.1\arcsec \citep{Combes2019}
\end{list}
\end{table*}

\section{Observations and data analysis}
\label{sec:obs}

The observations were carried out with the ALMA interferometer
in cycle 7, with 38 to 48 antennas, during september
2021. The corresponding ALMA project ID was
2019.1.00273.S with PI F. Combes.
Two galaxies (NGC 613 and NGC 1672) were observed simultaneously in
CO(3-2), HCO$^+$(4-3), and continuum, with Band 7 \citep{Mahieu2012}.
The extended configuration (baselines between 122~m to 16.2~km) used resulted in a
synthesized beam of 0.015-0.012\arcsec (0.8-1.2~pc for NGC~1672 and NGC~613 respectively), 
and an rms sensitivity of $\sim$0.4 mJy beam$^{-1}$ in 5.7 \kms\, 
channels (19 $\mu$Jy beam$^{-1}$ in the 3.7 GHz band
of the continuum). The maximum recoverable scale is 0.24\arcsec\, (13 and 20~pc 
for NGC~1672 and NGC~613 respectively).
Each galaxy was observed during 3 periods of about 42~min on source. 
 The total integration time, including calibration and
overheads, was 6 h per source. The choice of correlator configuration, 
selected to simultaneously observe two lines, provided a
velocity range of 1350 \kms\, for each line. 

\begin{figure*}
\begin{center}
\includegraphics[clip,width=0.48\textwidth,angle=0]{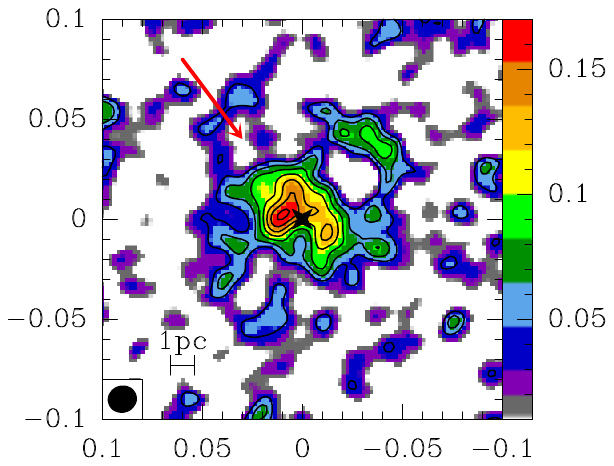}
\includegraphics[clip,width=0.46\textwidth,angle=0]{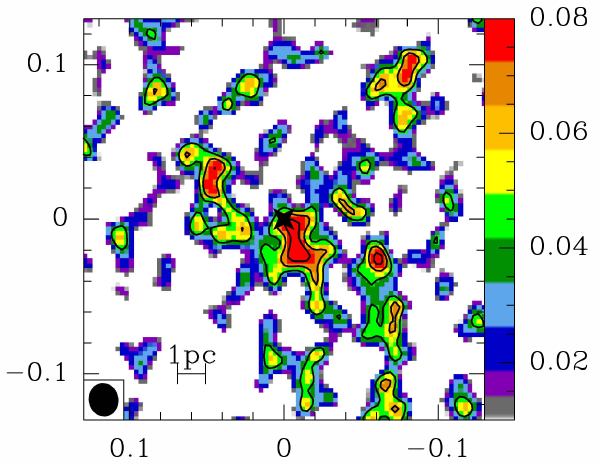}
\vskip+0.0cm
\caption{Continuum map at 0.86mm of the NGC~613 (left) and NGC~1672 (right) nucleus.
The spatial scale is RA-DEC  offset in arcsec from the central
position of Table \ref{tab:samp}. The color scale is in mJy beam$^{-1}$.
The first contour is at 2$\sigma$, and the next contours spaced by 1$\sigma$,
see Sec. \ref{sec:cont} for the value of $\sigma$.
The AGN center is indicated by a black cross, and the bar at bottom left is 1~pc long.
 A red arrow indicates the jet direction in NGC~613.} 
\label{fig:both-cont} 
\end{center}
\end{figure*}  

 The flux calibration was done with nearby quasars, which are regularly
monitored at ALMA, and resulted in a nominal 10\% accuracy.
The data were first calibrated with the CASA software, using the
scriptForPI and the
version 6.2.1.7 \citep{McMullin2007}. Then the analysis was finalized, 
through imaging and cleaning with the GILDAS software
\citep{Guilloteau2000}. The final cubes at high resolution are
at the maximum 6250 × 6250 pixels with 2.9~mas per pixel in the
plane of the sky, and have 236 channels of 6.77~MHz = 5.9~\kms\, width. 
Although these maps cover the FWHM primary beam of 18\arcsec, they show
signals only in the very center, due to the decreasing signal to noise
ratio with distance from the nucleus. The spectral windows spw19 (342~GHz) and 
spw21 (364~GHz) served to compute the continuum, and were subtracted in the UV plane from
the spw25 (the CO(3-2) line)  and spw27 (the HCO$^+$(4-3) line).
The CO(3-2) maps were made with Briggs weighting and a robustness parameter 
of 0.5, i.e., a trade-off between uniform and natural weighting. 
The HCO$^+$(4-3) maps, due to lower signal-to-noise ratio, were done
with natural weighting, resulting in a beam of 0.018-0.019\arcsec.
The data were cleaned using a mask made from the integrated
CO(3-2) map. The final maps were corrected for primary beam attenuation.
Because of missing short spacings, the interferometer does
not detect smooth and extended emission, with scales
 larger than $\sim$0.24\arcsec, in each
channel map. The scales between 0.24 and 3\arcsec were mapped
correctly with our previous configurations, and flux comparisons
to check for missing flux are reported in next Section.
If the large velocity gradients
prevent the line maps from being too much affected, 
 the missing-flux problem is more severely affecting 
continuum maps.

\begin{figure*}
\begin{center}
\includegraphics[clip,width=0.95\textwidth,angle=0]{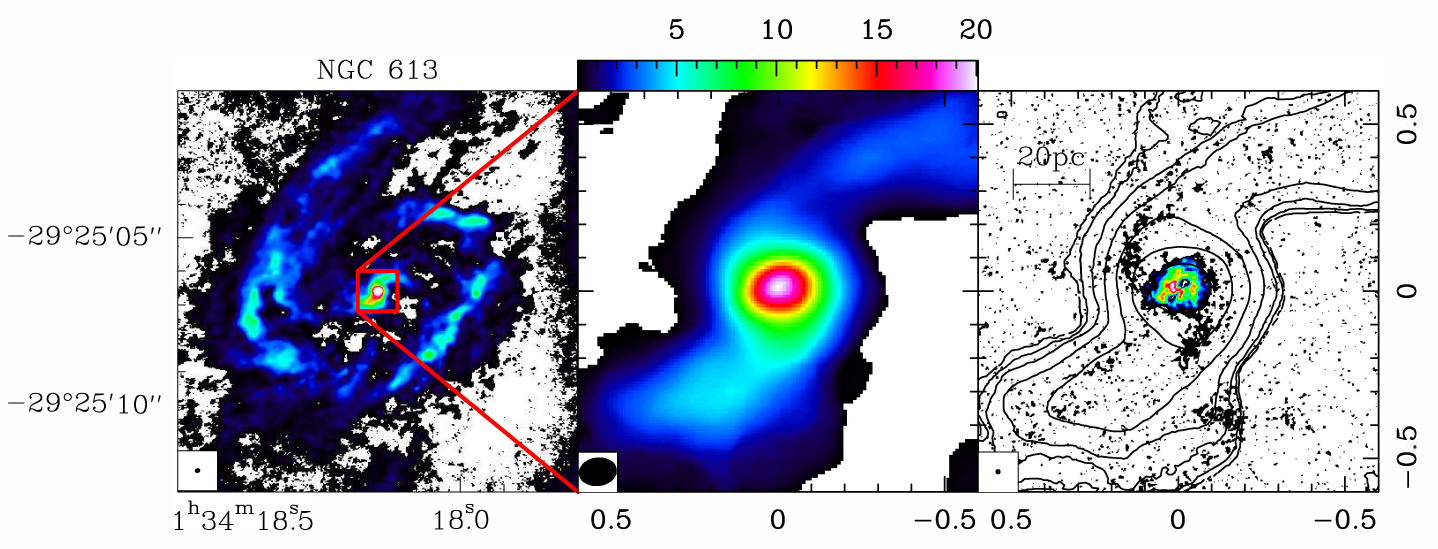}
\vskip+0.0cm
\caption{Low resolution (but without the most compact configurations) at $\sim$ 0.1'' (left), zoomed in the middle and high resolution at $\sim$ 0.015'' (right) of the CO(3-2) surface density toward NGC~613. The left and middle
moment-0 maps have been integrated over a velocity range -175 to 175~\kms, and with a 
threshold of 2.3~mJy beam$^{-1}$. In the right panel, the CO(3-2) contours of the low resolution map are overlaid.
While the left panel is plotted in absolute coordinates, 
the spatial scales of the two others are RA-DEC offsets in arc-seconds, from the center of Table \ref{tab:samp}.
The color-scale is in Jy beam$^{-1}$ \kms.} 
\label{fig:n613-lo-hires} 
\end{center}
\end{figure*} 

\begin{figure*}
\begin{center}
\includegraphics[clip,width=0.95\textwidth,angle=0]{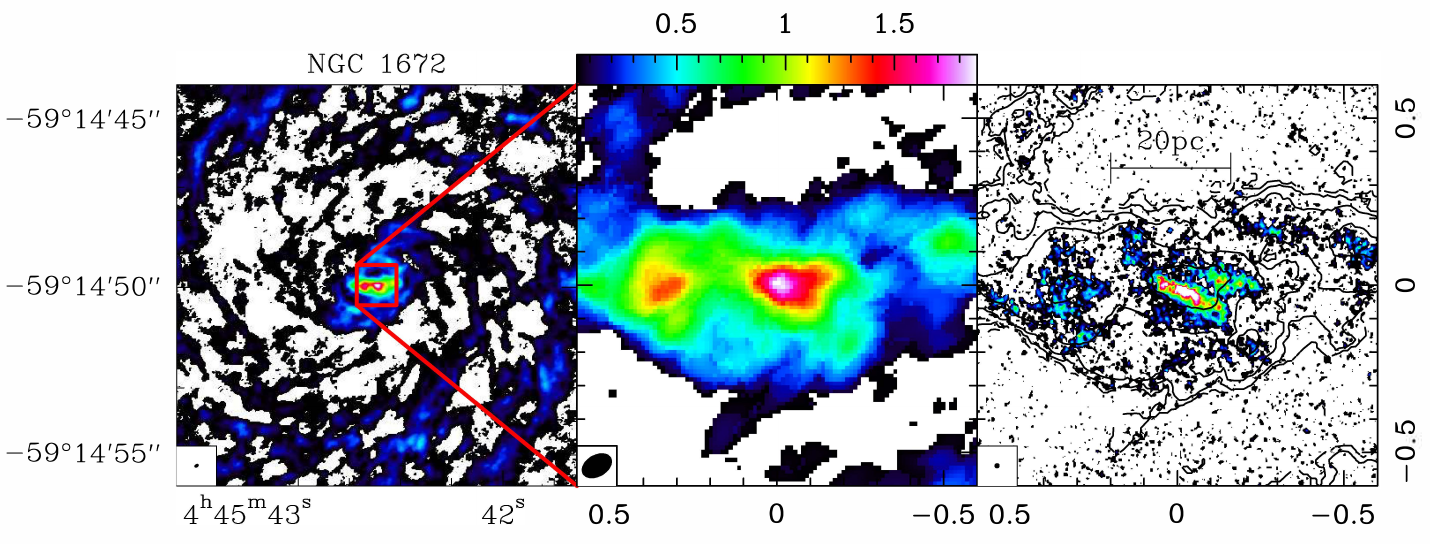}
\vskip+0.0cm
\caption{Same as Fig \ref{fig:n613-lo-hires}, for NGC~1672, with a different color bar. The left and middle
moment-0 maps have been integrated over a velocity range -200 to 200~\kms, and with a 
threshold of 4~mJy beam$^{-1}$. The color-scale is in Jy beam$^{-1}$ \kms.} 
\label{fig:n1672-lo-hires} 
\end{center}
\end{figure*} 
\section{Results}
\label{sec:res}

\subsection{Continuum maps}
\label{sec:cont}
 The continuum maps at 0.86~mm of the central regions of the two galaxies are displayed in
 Fig. \ref{fig:both-cont}. The signal is weak, since the long
 baselines filter out most of the signal. This means that the continuum emission is resolved.
 The continuum is weaker in NGC~1672, and the map was obtained in the natural weighting mode,
 while for NGC~613, we used the robust mode. The resulting beams are 14$\times$13 mas, PA=-72$^\circ$
 for NGC~613, and 22$\times$19 mas, PA=10$^\circ$  for NGC~1672. The peak continuum is 0.15~mJy beam$^{-1}$
 (7.5$\sigma$) and 0.10~mJy beam$^{-1}$ (5$\sigma$) respectively for the two galaxies. At low resolution,
 it was 2~mJy beam$^{-1}$, for a beam of 0.2\arcsec\, in NGC~613
 and 0.2~mJy beam$^{-1}$, for a beam of 0.1\arcsec\, in NGC~1672.
 Taking into account the different beam areas, it is difficult to estimate the exact flux lost,
 since the signal-to-noise does not allow us to map the wider beam, and the missing flux depends
 on the flux spatial distribution, which is extended for both resolution. It is clear however that
 the missing flux is higher in proportion than for the line (Sec. \ref{sec:morpho}).
 
 In the NGC~613 galaxy, \citet{Miyamoto2017} find in the center
 a radio jet with PA=20$^\circ$ at 95~GHZ, parallel to the VLA one at 4.9~GHz \citep{Hummel1992},
 but the nuclear emission is not resolved at 350~GHz with their beam of 0.44\arcsec. The nuclear
 continuum emission was partially resolved by \citet{Audibert2019} with the 0.2\arcsec\, beam and is extended
 in the SE direction, suggesting the association with the nuclear spiral.
 At a much smaller scale, with a total extension of 50~mas,
 We might see here the beginning of the jet, but with the somewhat 
 different position angle of PA=50$^\circ$.
 The weaker NGC~1672 continuum is also resolved, and might have an extension of 40~mas.
 
 It is not possible to disentangle the possible dust thermal contribution, 
 from the synchrotron emission, and for the latter the contribution of the radio jet and that
 of star formation. However, we may note that the continuum is much less extended than
 the molecular torus, suggesting that the thermal dust emission from the torus is not dominating.
 The continuum is centered at the same position as the AGN location, given in 
 Table \ref{tab:samp} within the beam of $\sim$ 0.1\arcsec \citep{Combes2019}.

 When taking into account higher frequency observations for CI and CO(6-5) (Audibert et al. in prep),
 it is possible to infer that the dust emission is not dominating in both galaxies.
 At 690~GHz, and a beam of 0.05$\times$0.04\arcsec, the continuum of NGC~1672 is not detected.
 In NGC~613, the continuum is detected at 690~GHz, and is 3~mJy beam$^{-1}$ with the same beam.
 However, it is not detected at 492~GHz ($<$ 0.5~mJy beam$^{-1}$), with this same beam, 
 suggesting that the synchrotron dominates.

\begin{figure}
\begin{center}
\includegraphics[clip,width=0.22\textwidth,angle=0]{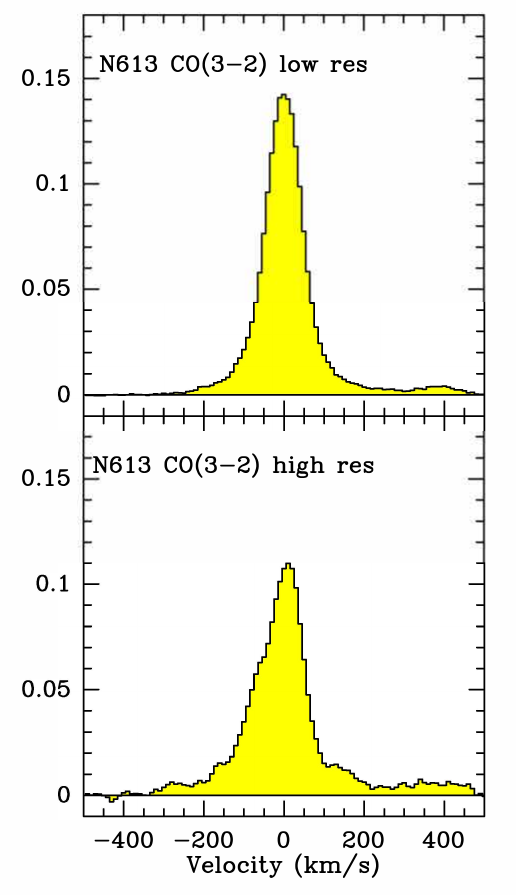}
\includegraphics[clip,width=0.22\textwidth,angle=0]{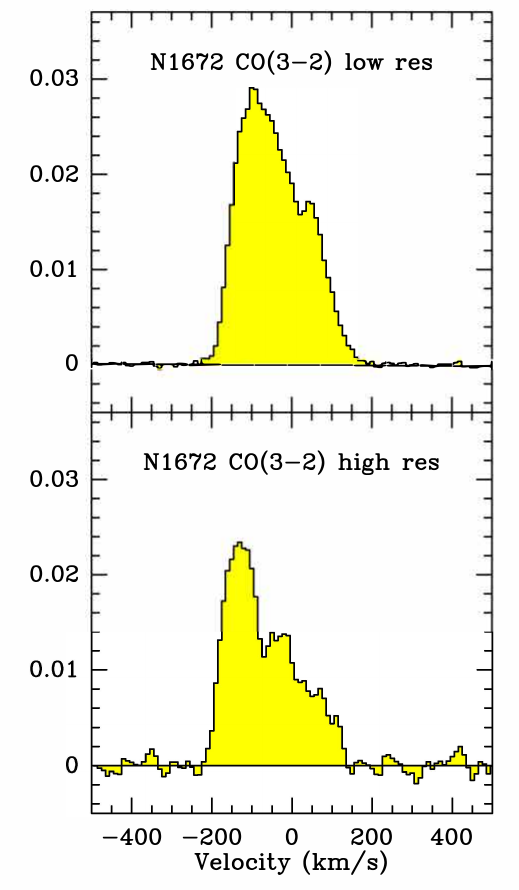}
\vskip+0.0cm
\caption{Comparison of the total spectrum within a circle of 0.2\arcsec\, in diameter between
the high resolution CO(3-2) observation of NGC~613 (left, bottom) and low resolution (left, top). 
Same for NGC~1672 at right. The vertical scale is Flux in Jy. 
 The involved aperture is much larger than the torus here.} 
\label{fig:both-spec2} 
\end{center}
\end{figure} 

\subsection{CO(3-2) line maps: morphology}
\label{sec:morpho}

\begin{figure}
\begin{center}
\includegraphics[clip,width=0.18\textwidth,angle=0]{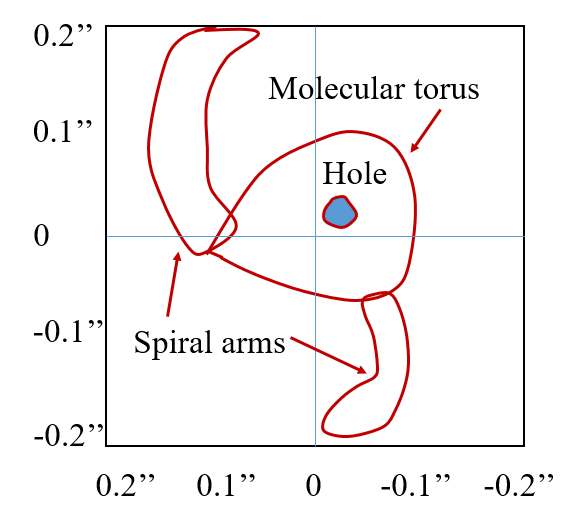}
\includegraphics[clip,width=0.30\textwidth,angle=0]{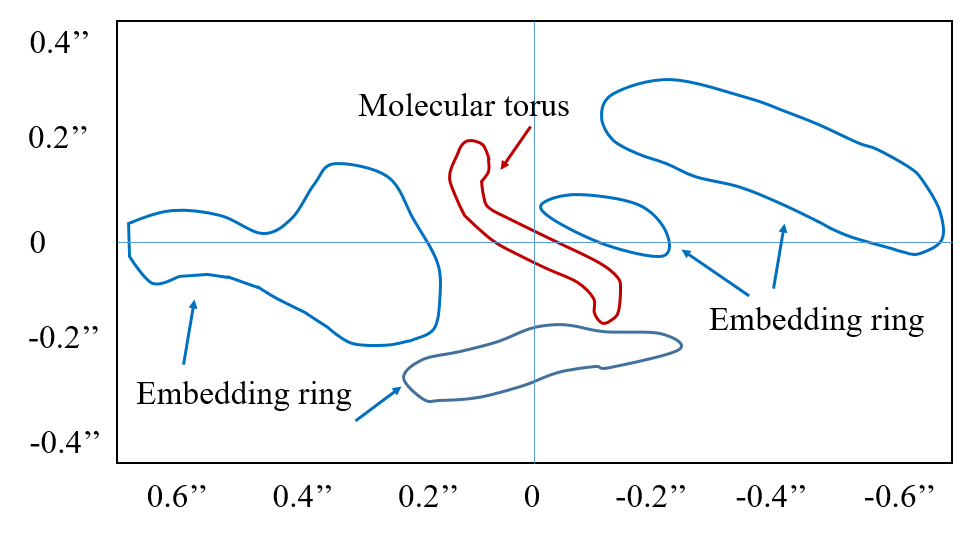}
\vskip+0.0cm
\caption{Cartoons to identify the principal components in NGC~613 (left) and NGC~1672 (right).  For
NGC~613, there is an H$_2$O maser detected in the (0,0) nucleus position \citep{Kondratko2006, Castangia2008}.
The scales are RA-DEC offsets in arc-seconds. The AGN centers are indicated by faint crosshairs.} 
\label{fig:both-cartoon} 
\end{center}
\end{figure}

Figures \ref{fig:n613-lo-hires} and \ref{fig:n1672-lo-hires} display the morphology of the molecular
gas at high resolution, with the zeroth-moment of the CO(3-2) line emission, compared with what was
obtained at the lower resolution of 0.2\arcsec, on the two galaxies
NGC~613 and NGC~1672. The beam at 
high resolution is $\sim$ 13 times smaller than that at low resolution, 
combining all previous observations with more compact configurations.
 The new maps unveil many unexpected details.
 In NGC~613, the molecular torus, inside the nuclear spiral, appears as a ring depleted in its center,
 more exactly  displaying a hole toward the NW direction, in both CO and HCO$^+$ lines. One possible interpretation is that the
 molecular torus is subject to an m=1 lopsided mode, and the present hole corresponds to a previous 
 strong AGN episode, which has ejected its surrounding gas. 
The torus is rather clumpy, and linked with two nascent spiral arms to the largest nuclear structure.
In NGC~1672, what was thought to be an i=66$^\circ$-inclined molecular torus, decoupled from 
the rest of the galaxy disk with PA=90$^\circ$, is now split in two components. The main nuclear molecular torus
is quite edge-on, with a PA=70$^\circ$, and is warped at its two extremities. The NE side 
is more extended, and its orientation tends toward the North axis.
The clumpiness of molecular tori has been modeled already to explain their spectral dust properties 
\citep[e.g.][]{Nenkova2008,Hoenig2010}.
Fig. \ref{fig:both-cartoon} display schematic cartoons to better identify the various
components in both galaxies.

 {\it Sizes:}  The high resolution enables us to distinguish the actual circum-nuclear components, that were not resolved before. In NGC~613, the true molecular torus has a radius of 0.1\arcsec\, = 8~pc, about twice smaller than previously determined \citep{Combes2019}. There is a depleted "hole" near the center of a beam size, i.e. a radius of 0.5~pc. The hole is offset toward the NW from the center, by 36~mas = 3~pc. This means some
 lopsidedness in the torus. The starting spiral arms span a total diameter of 50~pc, before merging into the nuclear spiral.
 In NGC~1672, the actual edge-on molecular torus has a radius of 0.13\arcsec\, = 7~pc, nearly four times smaller than previously identified. This is however the straight part of this edge-on structure, which
 extends through its NE warp to a radius of $\sim$ 14~pc. The molecular torus is embedded in a wider structure,
 kinematically decoupled, and more face-on, with a radius of 0.8\arcsec\, = 44~pc. This latter structure corresponds to the unresolved central region conspicuous in [FeII] and H$_2$(1-0)S(1), tracing the AGN excitation \citep{Fazeli2020}. This structure is then rather isolated at the center of the inner
 Lindblad resonance ring at a radius of 5\arcsec\, = 275~pc, which is actively forming stars
 \citep{Jenkins2011}.

 {\it Masses:} To derive the H$_2$ mass of observed components,
 we compute L$^\prime_{CO}$, the CO luminosity in  units of K \kms pc$^2$,
with the integrated emission in the beam. This CO luminosity is given by
\begin{equation}
L'_{CO} =
3.25 \hskip3pt 10^7 S_{CO}dV  {{D_L^2}\over {\nu_{rest}^2(1+z)}} \hskip6pt \rm{K\hskip3pt
  km \hskip3pt s^{-1}\hskip3pt pc^2},
\end{equation}
where  $S_{CO}dV$ is the integrated flux in Jy \kms\,, $\nu_{rest}$ the rest
frequency in GHz,  and $D_L$ the luminosity distance in Mpc.
Under the  assumption of a standard CO-to-H$_2$ conversion factor \citep{Bolatto2013},
we compute the H$_2$ mass using M(H$_2$) = $\alpha$ L'$_{\rm CO}$,
with $\alpha=4.36$ M$_\odot$ (K \kms\, pc$^2$)$^{-1}$.  
This relation is calibrated for the fundamental CO(1-0)
line, but should also be valid for any higher $J$-level, provided LTE excitation at high
kinetic temperatures. However, the higher levels are sun-thermally excited in galaxies, and
 we adopted the $R_{1J} = T_1/T_J$ empirical correction ratios from 
\citet[e.g.][]{Tacconi2018}, that is for CO(3-2) $R_{13} = 1.8$.  The molecular gas mass estimates for the two galaxies are presented in Table~\ref{tab:spec2} and Table~\ref{tab:quant}. 

The maximum observed H$_2$ column densities derived in Table~\ref{tab:quant} imply
that the two nuclei are Compton-thick. This means that the strongest X-ray sources, observed
toward the nuclei, can be due to the active AGN, while there was some debate around
the nature of these sources in NGC~613 \citep{Castangia2013} and in
NGC~1672 \citep{deNaray2000,Jenkins2011}.

\begin{table}
\caption{Comparison of low and high resolution CO(3-2) central 0.2\arcsec\, fluxes. Given are the 
results of the Gaussian fits.}
\vspace{-0.4cm}
\begin{center}
\begin{tabular}{rccccc}
\hline  \hline
Galaxy  & Area   & FWHM  &  S$_{peak}$ & M(H$_2$)\\
        &  (Jy \kms) &  \kms &  (mJy) & 10$^6$ M$_\odot$\\
   \hline
   N613-low   & 16.8$\pm$0.14  & 112$\pm$1   & 140$\pm$3    &10.6$\pm$0.1\\
      -high   & 15.4$\pm$0.28  & 144$\pm$3   & 100$\pm$5    &9.7$\pm$0.2 \\
   N1672-low  & 5.91$\pm$0.09  & 202$\pm$0.1 & 27.5$\pm$0.2 &1.6$\pm$0.02 \\
       -high  & 4.14$\pm$0.06  & 198$\pm$3   & 19.6$\pm$0.8 &1.1$\pm$0.16\\
\hline 
\end{tabular}
\tablefoot{The low and high resolution are 0.2 and 0.015\arcsec\, respectively}    
\label{tab:spec2}
\end{center}
\end{table}

To quantify whether we are missing flux in the central region of the galaxies,
we compare the central beam spectrum of the previous maps, all configurations combined,
resulting in a beam of
0.2\arcsec. We integrate the new high resolution CO(3-2) maps over such a region,
and the comparisons are plotted in Fig. \ref{fig:both-spec2}
for NGC~613 (left panel) and NGC~1672 (right panel) respectively. We conclude from the quantification of 
Table~\ref{tab:spec2} resulting from Gaussian fits, that the missing flux is variable, according to the concentration
of the molecular gas. From the integrated areas it is 9\% for NGC~613 and 45\% for NGC~1672.

\begin{figure*}
\begin{center}
\includegraphics[clip,width=0.99\textwidth,angle=0]{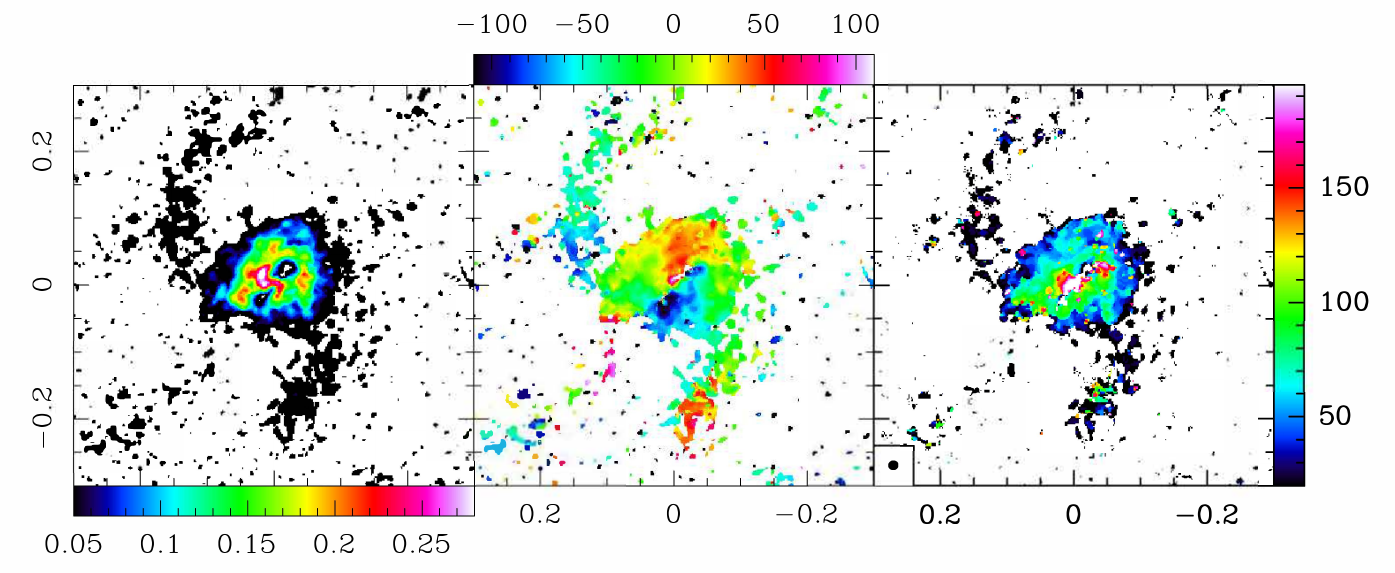}
\vskip+0.0cm
\caption{Moment zero (left), one (middle) and two (right) of the CO(3-2) cube with 0.015\arcsec beam toward NGC~613.
 An H$_2$O maser has been detected coinciding with the (0,0) position \citep{Kondratko2006, Castangia2008}.
Color scale to the left is in Jy beam$^{-1}$ \kms.
The moments have been integrated over a velocity range -300 to 300~\kms, and with a 
threshold of 0.6~mJy beam$^{-1}$.
In the middle, the color scale is in velocity difference from the V$_{sys}$= 1481 \kms.
On the right the color-scale is given in \kms.
The synthesized beam is shown in the bottom-left corner of right box.
The spatial scales are RA-DEC in arc-seconds, from the central position reported in Table \ref{tab:samp}.} 
\label{fig:n613-mom} 
\end{center}
\end{figure*} 

\begin{figure*}
\begin{center}
\includegraphics[clip,width=0.99\textwidth,angle=0]{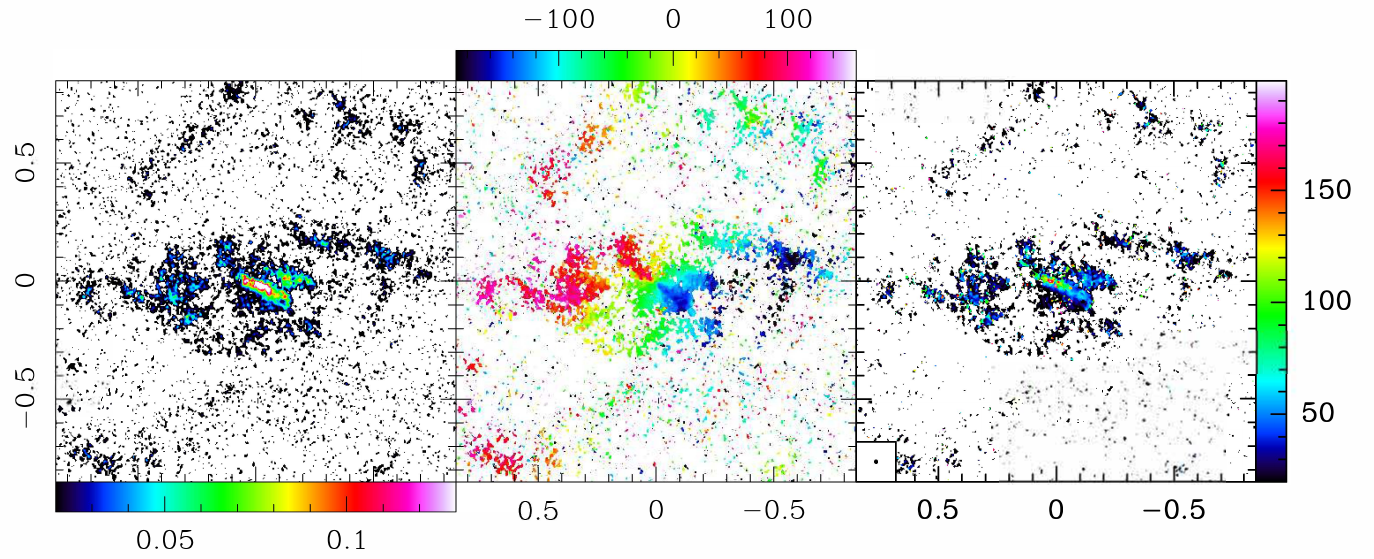}
\vskip+0.0cm
\caption{Same as Fig \ref{fig:n613-mom}, for NGC~1672.  
The moments have been integrated over a velocity range -250 to 250~\kms, and with a 
threshold of 0.55~mJy beam$^{-1}$. In the middle, the color scale is in 
velocity difference from the V$_{sys}$= 1331 \kms.} 
\label{fig:n1672-mom} 
\end{center}
\end{figure*}

\subsection{CO(3-2) line: Kinematics}
\label{sec:kine}

Fig.~\ref{fig:n613-mom} and Fig.~\ref{fig:n1672-mom} show the three first moments of 
the NGC~613 and NGC~1672 cubes.
For the NGC~613 galaxy, the velocity field now shows a clear decoupling from the host disk,
with a kinematic position angle of PA = -20$^\circ$. This PA corresponds also to the elongated
shape of the zeroth-moment, which appears like an ellipse, somewhat truncated
at the SE extremity. From this oval shape, an assuming a circular torus, 
an inclination of 50$^\circ$ can be derived. There
is a strong velocity gradient toward the center, and this is reflected in the second moment,
where the velocity dispersion appears to maximize near the major-axis, up to 300~\kms. 

For the NGC~1672 galaxy, the central velocity field is more complex since there is
a superposition of two components, with different inclinations on the line of sight. 
However the velocity gradient of the edge-on molecular torus is very clear, and quite
decoupled from the other more face-on (i= 68$^\circ$) component.  The molecular torus
has a kinematical PA = 70$^\circ$, more perpendicular to the large-scale PA= 124$^\circ$ 
of the galaxy disk \citep{Diaz1999}. Since the galaxy is less distant,
the beam can better resolve the strong velocity gradient toward the center, and the
apparent velocity dispersion in the second moment map maximize at only 200~\kms.

\subsubsection{Molecular gas outflow}
\label{sec:outflow}

\begin{figure}
\begin{center}
\includegraphics[clip,width=0.2\textwidth,angle=0]{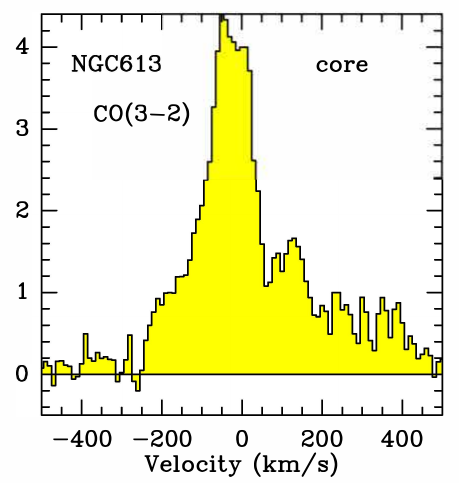}
\includegraphics[clip,width=0.22\textwidth,angle=0]{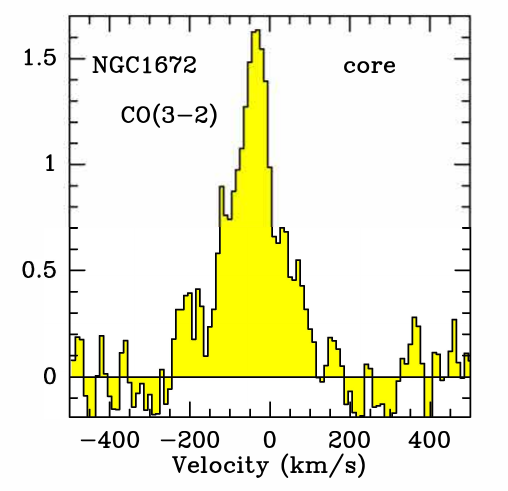}
\vskip+0.0cm
\caption{ CO(3-2) average spectrum from the high resolution configuration, of the 
0.014\arcsec\, central beam of NGC~613, showing high velocity components (left),
and of NGC~1672 (right). The vertical scale is Flux in mJy.} 
\label{fig:both-spec1} 
\end{center}
\end{figure}

\begin{figure}
\begin{center}
\includegraphics[clip,width=0.232\textwidth,angle=0]{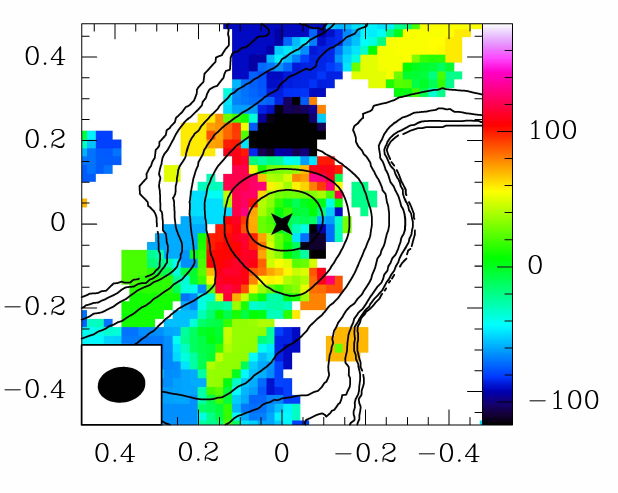}
\includegraphics[clip,width=0.252\textwidth,angle=0]{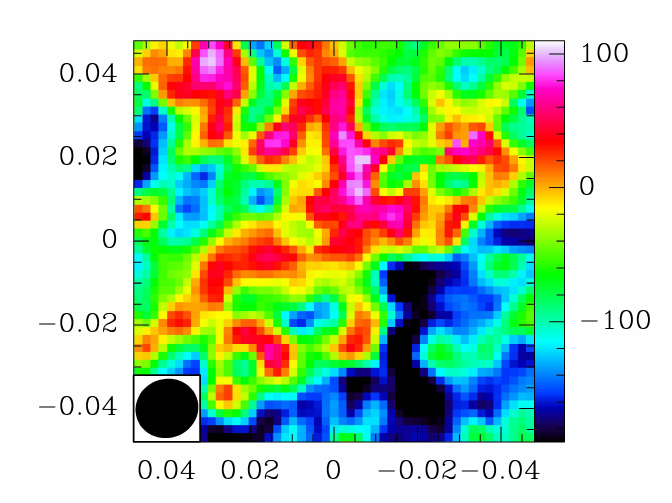}
\vskip+0.0cm
\caption{Velocity map of the NGC~613 CO(3-2) line residuals, once the spectra have been fitted by a Gaussian,
inside the range -150 to 150 \kms, first at low resolution (left) and then
at high resolution (right). Note that the field of view is different in the two panels
(1\arcsec, and 0.1\arcsec, respectively).
The contours of the CO(3-2) map at low resolution are
overlaid, in the left panel. The scale is RA-DEC offset in \arcsec. 
 The color scale is in \kms.} 
\label{fig:n613-outflow} 
\end{center}
\end{figure}

A molecular outflow was discovered at low spatial
resolution in NGC~613 \citep{Audibert2019}. The molecular outflow mass 
was M$_{out}$ = 2$\times$10$^6$ M$_\odot$ and the mass outflow rate 
27~M$_\odot$/yr. The outflow was observed in the same
direction as the radio jet mapped with the VLA, suggesting that the gas was
entrained by the jet. With the present higher resolution observations, we detect
clearly broad wings on each side of the central spectrum, in Fig. \ref{fig:both-spec1}.
The total observed spectrum extends to 650~\kms\, on each side and does not show
any more signal beyond $\pm$500~\kms.
   However, the red wing is contaminated by the H$^{13}$CN(4-3) line, at 345.340~GHz,
 situated at 396~\kms from the CO(3-2) line. The spectrum beyond 300~\kms\, has therefore to
be ignored. 
   In order to quantify the molecular outflow in the high resolution map, we extracted 
the central 0.15\arcsec$\times$0.15\arcsec region, corresponding to
75$\times$75 = 5625 pixels (2~mas per pixel), between velocity -500 to 300~\kms, and 
fitted the main spectral component between -150 and 150~\kms\, by a Gaussian. Removing
the Gaussian fit from every pixel, we mapped the residual emission, where the maximum
integrated flux was 10 times lower than the original map. The velocity field of 
 this residual map is still indicating red shift in the North and blue shift in the South,
 see Fig. \ref{fig:n613-outflow}, right panel.
   There is no evidence of the outflow, in this high resolution map. The previously
identified outflow shows a blue wing in the NE and a red wing in the SW. 
 The outflow velocities are opposite to the rotation velocities, with also a different
 position angle, which helps to reduce the confusion between the two components.
 
\begin{figure}
\begin{center}
\includegraphics[clip,width=0.5\textwidth,angle=0]{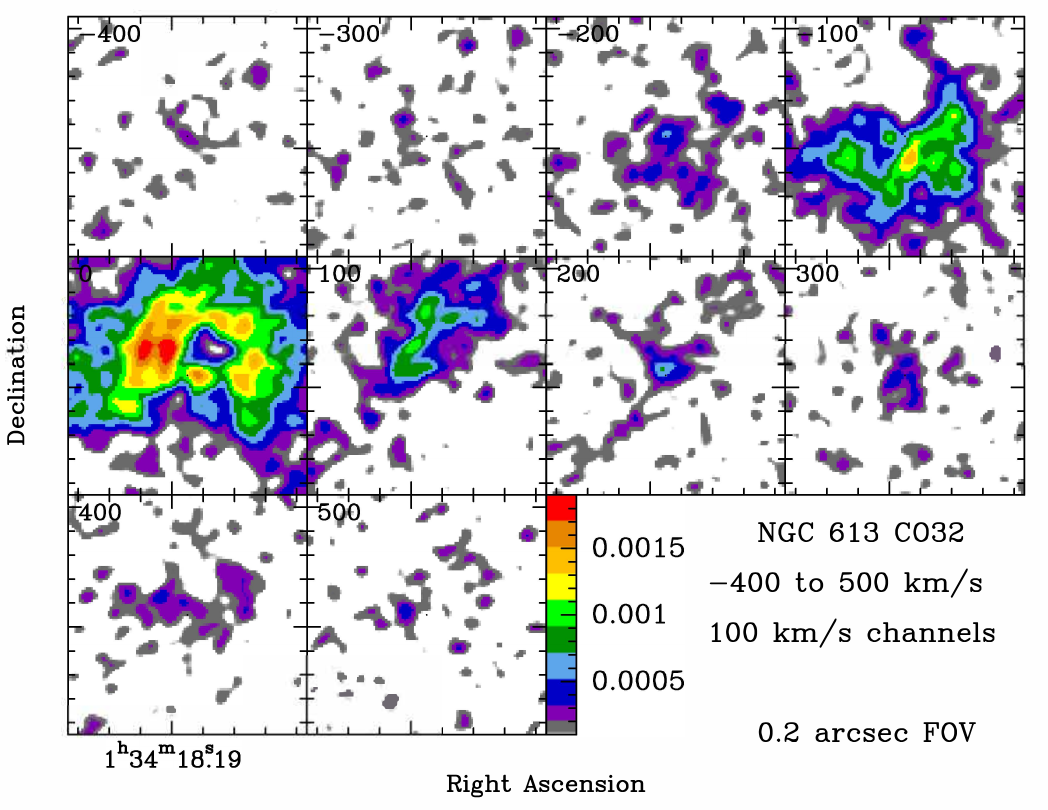}
\vskip+0.0cm
\caption{Channel map in CO(3-2) of NGC~613.  Each channel is 100~\kms\, wide, to show the high-velocity component 
with the same flux scale. The latter is in Jy beam$^{-1}$, average over the channel velocity range. Each panel is 
 0.2\arcsec in size.} 
\label{fig:n613-channels} 
\end{center}
\end{figure}

The channel map of Fig. \ref{fig:n613-channels} confirms the velocity extension in particular
on the redshifted side, even taking into account the  H$^{13}$CN(4-3) contamination in the two last
positive channels. The absence of the outflow signature at these small scales may be due to 
the interferometer filtering the more extended features. To check this, we applied the same 
gauss fit technique as described above to the more extended CO(3-2) cube with all configurations
included. The derived residual map reveals now the outflow, as shown in Fig. 
\ref{fig:n613-outflow}, left panel.
This is consistent with the outflow derived previously by \citet{Audibert2019}.
The outflow is seen at 25~pc scale, which is larger than the maximum recoverable 
scale of 20pc with the high-resolution configuration. 
The broad wings of NGC~613 central spectra, now dominated by the rotation around
the black hole (see Sec. \ref{sec:BHmass}), were not dominant at low resolution, when 
the emission was dominated by gas at much larger scale.

The disappearing of the outflow at high resolution was also seen in NGC~1068
 \citep{GRAVITY2020,Gamez-Rosas2025}. This may be a generic property of these winds. If they are resolved out,
 this means that the wind is quite diffuse, with little small-scale structure.
 Alternatively, the multi-phase outflow could be ionized only near the center 
 \citep[e.g.][]{Alonso-Herrero2019,Garcia-Bernete2021}.

The outflow recovered in Fig. \ref{fig:n613-outflow} is also consistent
in orientation with the much larger scale outflow 
(10\arcsec\, or 1~kpc size) detected in the [OIII] line by \citet{Hummel1987}. 
The excitation of the ionized gas in the outflow is mixed from both star formation and AGN \citep{Silva-Lima2025}, and the outflow kinetic energy could also be due to the combination of several parameters
acting together, AGN winds and jets or massive stars feedback.

There is no evidence of outflow in the center of NGC~1672. However, there is a radio source, unresolved with the 3cm beam of ATCA of 1.3\arcsec\, \citep{Jenkins2011}, associated with the AGN, that could correspond
to a radio jet. There are no broad wings localized in the nucleus, as in NGC~613, may be because 
the molecular gas is not as much concentrated, see Fig. \ref{fig:n1672-channels}.  This will be
obvious in the next section, in the position-velocity diagram.

\subsubsection{Black hole mass}
\label{sec:BHmass}

With the high resolution of 1~pc scale, our measurements of velocities are
well within the sphere of influence (SoI) of the black holes, whatever their 
definition. A first definition SoI1, according to \citet{Merritt2004} is
the radius where the enclosed mass is twice the black hole mass M$_{BH}$.
Another definition is SoI2 = G M$_{BH}$/$\sigma_v^2$, where
$\sigma_v$ is the central velocity dispersion of the stars,
essentially the central bulge velocity dispersion. These values are
11-50~pc and 33-8.8~pc for NGC~613 and NGC~1672 respectively.
Within 10~pc, the bulge mass has been computed by
\citep{Combes2019} from a mass model
based on HST and Spitzer observations to be 2.7 and 4.1 $\times$ 10$^6$ M$_\odot$
for NGC~613 and NGC~1672 respectively. This means that we can get
a good approximation of M$_{BH}$ by computing the enclosed mass,
and subtracting these values.

\begin{figure}
\begin{center}
\includegraphics[clip,width=0.48\textwidth,angle=0]{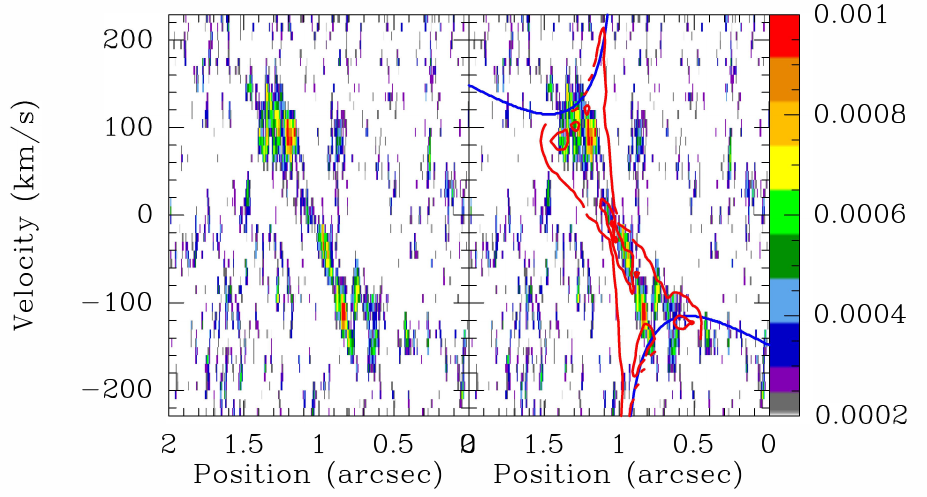}
\vskip+0.0cm
\caption{Position-velocity diagram in CO(3-2) for NGC~1672, taken along the  major axis of the torus,
PA = 70 $^\circ$. The two panels are the same, with at right the superposition in blue of the
circular velocity of the galaxy model with M$_{BH}$ = 4.5 $\times$ 10$^7$ M$_\odot$. The red contours are from
the gas density of the torus model. The strip of length 2\arcsec\, is centered on the galaxy at 1\arcsec.
The color bar scale is Jy beam$^{-1}$.} 
\label{fig:n1672-pv} 
\end{center}
\end{figure}

The velocity gradient in the edge-on torus of NGC~1672 can be measured from
the position-velocity diagrams in Fig. \ref{fig:n1672-pv}. There is a rotational velocity
of 140~\kms\, at a radius of 0.2 \arcsec\, = 11~pc.  For gas in a circular orbit, this means
an enclosed mass of 5 $\times$ 10$^7$ M$_\odot$.  The bulge mass inside 11~pc is
4.9 $\times$ 10$^6$ M$_\odot$, and we can derive M$_{BH}$ = 4.5 $\times$ 10$^7$ M$_\odot$.
This value is consistent with previous determinations \citep[e.g.][]{Combes2019},
well within the error bars. It is significantly larger than that derived from the M-$\sigma$ relation
of 2.5 $\times$ 10$^7$ M$_\odot$.
This might be surprising, since we observe a comparable velocity gradient, but over a smaller 
distance, thanks to the higher resolution. However, in the previous model, we corrected the velocity
for the inclination, and now the torus is edge-on. We also computed the contours of the
position-velocity diagram, as done in \citep{Combes2019}, 
assuming circular orbits. These are displayed in the right panel of Fig. \ref{fig:n1672-pv}.

\begin{figure}
\begin{center}
\includegraphics[clip,width=0.48\textwidth,angle=0]{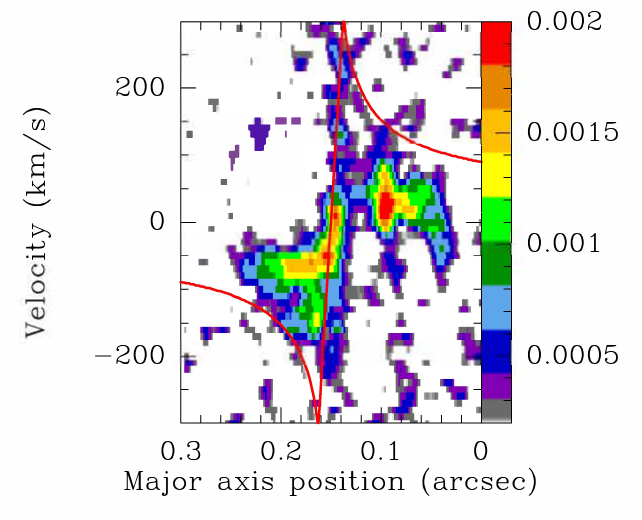}
\vskip+0.0cm
\caption{Position-velocity diagram in CO(3-2) for NGC~613, taken along the  major axis of the torus,
PA = -20 $^\circ$. 
 The circular velocity predicted in projection for M$_{BH}$ = 3.4$\times$ 10$^7$ M$_\odot$ is superposed in red.
 The strip of length 0.3\arcsec\,is centered on the galaxy at 0.15\arcsec. The color bar scale is Jy beam$^{-1}$.} 
\label{fig:n613-pv} 
\end{center}
\end{figure}

For NGC~613, the orientation of the torus of i=50$^\circ$ allows us to ascertain
that assuming circular orbits is a reasonable approximation. On the kinematic axis of
PA= 160$^\circ$, we can measure a projected rotation velocity of 250~\kms\, at 1~pc radius
in  Fig. \ref{fig:n613-pv}. After correcting for the inclination, this leads
to an enclosed mass of 3.4 $\times$ 10$^7$ M$_\odot$. While the bulge mass enclosed
is estimated at 2.6$\times$  10$^4$ M$_\odot$, M$_{BH}$ = 3.4 $\times$ 10$^7$ M$_\odot$.
This leads to a value perfectly consistent with previous estimates, 
and with that from the M-$\sigma$ relation of 3.7 $\times$ 10$^7$ M$_\odot$ 
\citep{Combes2019}.
 Finally, the lack of resolution was nicely compensated by a correct modeling
 of the contribution of the other mass components, while at high resolution,
 the latter are not important.  The black hole masses are now more robust.

\begin{table}
\caption{Derived quantities for the molecular tori of the two galaxies}
\vspace{-0.4cm}
\begin{center}
\begin{tabular}{lccccc}
\hline  \hline
Galaxy  & Radius & M(H$_2$)     &  inc    & PA     & logN(H$_2$) \\
        &  pc  & 10$^7$ M$_\odot$&$^\circ$&$^\circ$&  cm$^{-2}$   \\
   \hline
   N613   & 8    & 1.0      &   50      &   -20  &    24.9   \\
   N1672  & 14   & 0.2      &   90      &    70  &    24.6   \\
\hline 
\end{tabular}
\label{tab:quant}
\end{center}
\end{table}

\begin{figure}
\begin{center}
\includegraphics[clip,width=0.22\textwidth,angle=0]{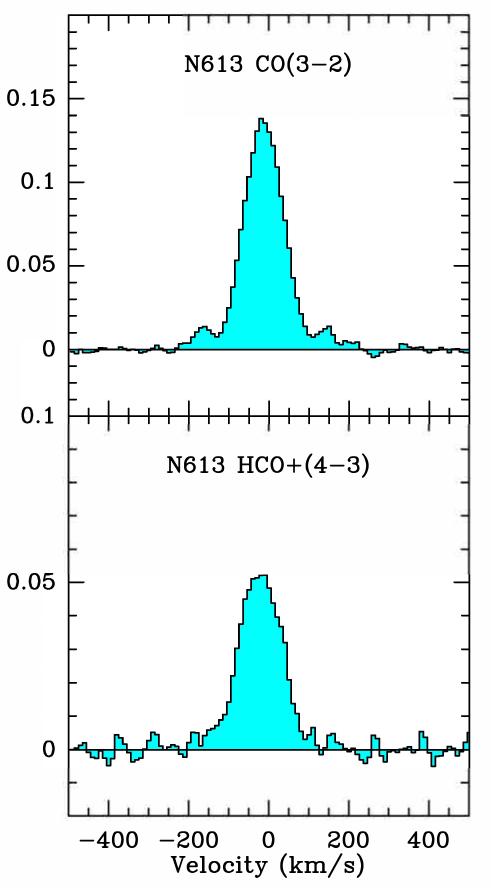}
\includegraphics[clip,width=0.22\textwidth,angle=0]{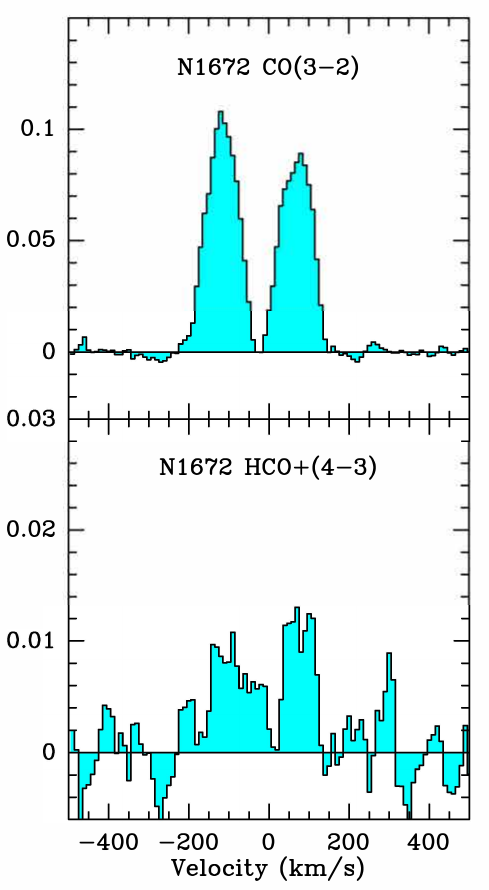}
\vskip+0.0cm
\caption{Total spectra in the high resolution configuration, integrated over the field
of view containing emission, 0.6\arcsec\, for NGC~613 (left) and 1.7\arcsec\, for 
NGC~1672 (right), in CO(3-2) (top), and HCO$^+$(4-3) (bottom).  
The vertical scale is Flux in Jy.}
\label{fig:both-spectot} 
\end{center}
\end{figure} 


\subsection{HCO$^+$(4-3) line}
\label{sec:HCO}

\begin{figure*}
\begin{center}
\includegraphics[clip,width=0.9\textwidth,angle=0]{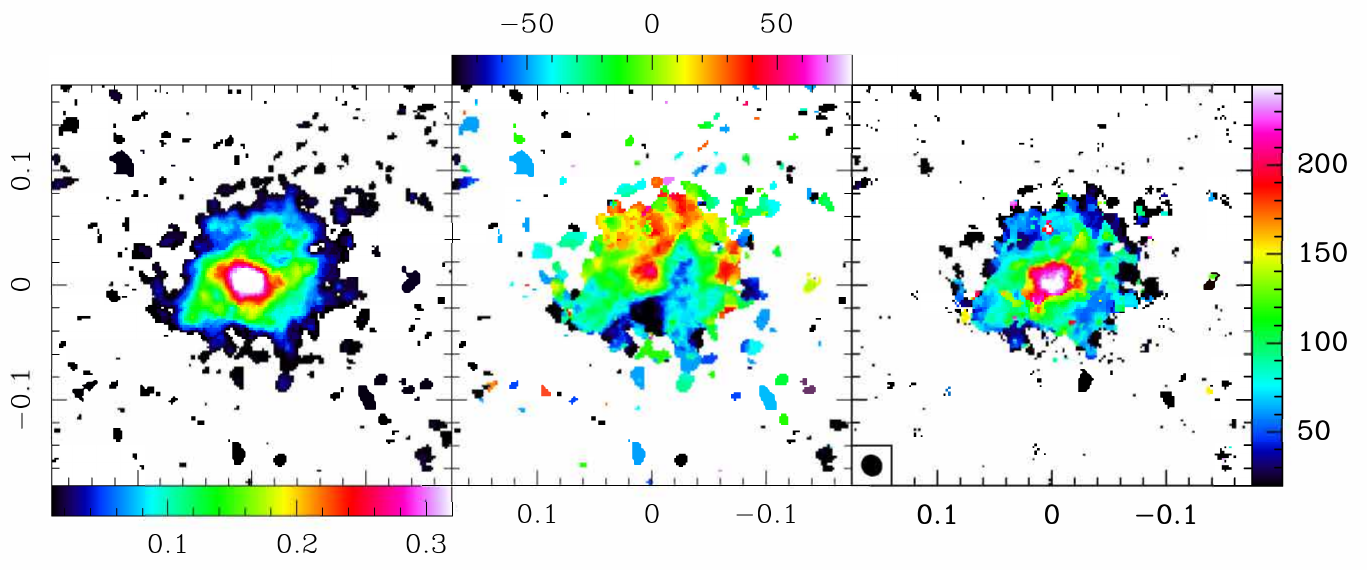}
\vskip+0.0cm
\caption{Moment zero (left), one (middle) and two (right) of the HCO(4-3) cube with 0.019\arcsec\, beam toward NGC~613. The moments have been integrated over a velocity range -260 to 260~\kms, and with a 
threshold of 2.2~mJy beam$^{-1}$.
In the middle, the color scale is in velocity difference from the V$_{sys}$= 1481 \kms.
The synthesized beam is shown in the bottom-left corner of right box.
The spatial scales are RA-DEC in arc-seconds, from the central position reported in Table \ref{tab:samp}.} 
\label{fig:n613-mom-hco} 
\end{center}
\end{figure*}

\begin{figure*}
\begin{center}
\includegraphics[clip,width=0.9\textwidth,angle=0]{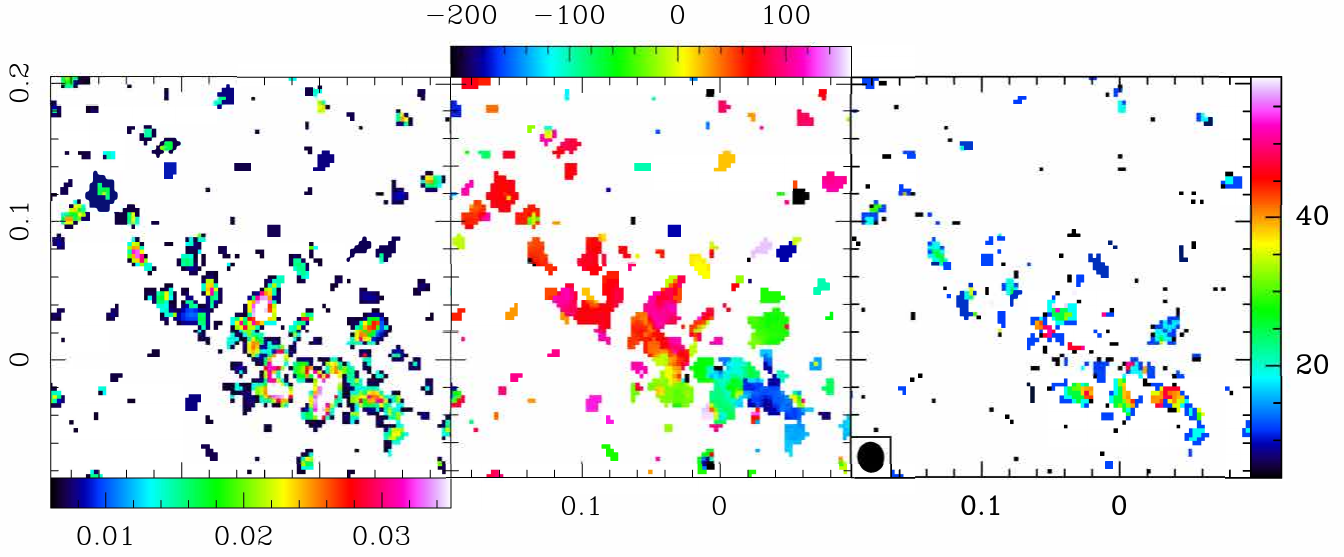}
\vskip+0.0cm
\caption{Same as Fig \ref{fig:n613-mom-hco}, for NGC~1672.
The moments have been integrated over a velocity range -260 to 260~\kms, and with a 
threshold of 0.65~mJy beam$^{-1}$. In the middle, the color scale is in 
velocity difference from the V$_{sys}$= 1331 \kms.} 
\label{fig:n1672-mom-hco} 
\end{center}
\end{figure*}

One of the spectral windows was dedicated to the HCO$^+$(4-3) line, a dense gas tracer,
yielding some new information, although its intensity is $\sim$ 3-6 times lower than that of CO(3-2).
The critical density to excite HCO$^+$(4-3) is 1.8 $\times$ 10$^6$cm$^{-3}$ \citep{Greve2009}.
Figures \ref{fig:n613-mom-hco} and \ref{fig:n1672-mom-hco} display the three first moments
of the HCO$^+$(4-3) map in the galaxies NGC~613 and NGC~1672 respectively.

In the galaxy NGC~613, the molecular torus is very concentrated in the dense gas tracer,
even more than in the CO(3-2) emission. The maximum in the HCO$^+$(4-3) zeroth-moment map
is in the core, with a value 0.47 Jy beam$^{-1}$ \kms,
while in CO(3-2), the maximum value is a bit offset, with 0.32 Jy beam$^{-1}$ \kms. 
The depleted region at 3~pc in the NW direction is still there, may be less
pronounced than in CO(3-2), because of the larger beam of 0.019\arcsec.  Taking into
account that the beam size has an area 1.65 larger in the HCO$^+$(4-3) map, the CO(3-2) maximum is 
still higher by a factor 1.1, but this ratio is surprisingly low, with respect to the common
ratios $\sim$ 3-6 quoted above.

In NGC~1672, the concentration of dense molecular gas is less pronounced. While the maximum value
of the CO(3-2) zeroth-moment map is 0.21 Jy beam$^{-1}$ \kms, the HCO$^+$(4-3) maximum is
only 0.072 Jy beam$^{-1}$ \kms, and given the beam ratios of 1.49, the CO(3-2) is 
stronger by a factor 4.3. 

To quantify the global emissions detected in the molecular torus and circum-nuclear region
of 50~pc in diameter, we plot in Fig. \ref{fig:both-spectot}, the total spectrum in 
CO(3-2) and HCO$^+$(4-3) for both galaxies. In Table \ref{tab:spectot}, the results of the Gaussian fits
for these total spectra are displayed. It can be seen that the global integrated signal ratio
between CO(3-2) and HCO$^+$(4-3) is 2.4 for NGC~613 and 8 for NGC~1672. It is therefore clear that
the ratio increases with distance from center, as expected, since the molecular gas is
much denser near the core, and this effect is even exacerbated in NGC~613. 
Another comparison can be done with the larger-scales sampled at low resolution
\citep{Audibert2019}. In NGC~613, the CO(3-2) to HCO$^+$(4-3) ratio is 40. It is the same order
of magnitude in NGC~1672, with more uncertainty. Such ratios are common at kpc scales in galaxies
\citep[e.g.][]{Usero2015,Izumi2016}.

The three moments for the HCO$^+$(4-3) line are quite similar to that of the CO(3-2), when taken
into account the lower sensitivity, in particular far from the nuclei. The low signal-to-noise ratio is 
more severe for NGC~1672, where the molecular torus is quite noisy. However, the strong velocity gradient 
is clear, and the warp of the torus to the NE as well.
The comparison between low and high resolution velocity fields are compared in
Fig. \ref{fig:n613-lo-hires-hco} and  \ref{fig:n1672-lo-hires-hco}.

\begin{table}
\caption{Comparison of total integrated fluxes in the central 50~pc circum-nuclear
regions of both galaxies, in the CO(3-2) and HCO$^+$(4-3) lines.}
\vspace{-0.4cm}
\begin{center}
\begin{tabular}{rccccc}
\hline  \hline
Galaxy  & Area   & FWHM  &  S$_{peak}$ & M(H$_2$)\\
        &  (Jy \kms) &  \kms &  (mJy) & 10$^7$ M$_\odot$\\
   \hline
   N613-CO32   & 17$\pm$0.2    & 116$\pm$1   & 137$\pm$3    &1.07$\pm$0.01\\
      -HCO43   & 7.2$\pm$0.1   & 125$\pm$3   &  54$\pm$2    &    \\
   N1672-CO32  & 10.5$\pm$0.09  & 90$\pm$1    & 109$\pm$2   &0.29$\pm$0.002\\
        -CO32  &  8.9$\pm$0.09  & 90$\pm$1    &  93$\pm$2   &0.25$\pm$0.002\\
       -HCO43  & 1.36$\pm$0.19  & 140$\pm$22  & 9.1$\pm$2   &    \\
      -HCO43   & 1.03$\pm$0.13  &  74$\pm$9   &  13$\pm$2   &    \\
\hline 
\end{tabular}
\tablefoot{The Gaussian fits for NGC~1672 were done on two components, corresponding to the two lines, for each molecule. Their velocities are -116.3 and 69.5 $\pm$0.4 \kms\, for CO(3-2), and 
-97.0 and 76.6 $\pm$6 \kms\, for HCO$^+$(4-3).}    
\label{tab:spectot}
\end{center}
\end{table}

\section{Discussion}
\label{sec:disc}

\subsection{NGC~613 model in view of the new morphology}
From low-resolution maps, the molecular torus was identified as
the nuclear disk of radius 14~pc, located inside the nuclear spiral,
and kinematically decoupled from the galaxy disk.
The high resolution map has kept nearly the same morphology and orientation, but reduced
the torus radius from 14 to 8~pc, and reduced its mass by more than a factor 2, from
3.9 to 1.4$\times$10$^7$ M$_\odot$, even taking into account the 
10\% missing flux. The torus is resolved in a lopsided ring.
The inclination and orientation of the torus might still be different
from those of the accretion disk and its surrounding inside it. The radio jet at cm wavelength
is oriented along the minor axis of the 300~pc-radius ring, unrelated to the torus.
 The velocity profile of the H$_2$O maser detected
close to the nucleus has a two-horn shape, with two narrow features, each
located at 40~\kms\, away from the systemic velocity \citep{Castangia2008}.
This indicates a more face-on orientation. 

\subsection{NGC~1672 model in view of the new morphology}
One of the striking features is the discovery at high resolution of the actual torus
which is edge-on and geometrically very thin. At low resolution it was embedded in another
component, making the ensemble appear at a lower inclination of 68$^\circ$, and with another
position angle (90$^\circ$). 
This structure surrounding the edge-on thin torus is a patchy ring extending 
from 0.3" to 0.5". Its radius of $\sim$ 22~pc may correspond to a
dynamical feature. NGC~1672 is a barred galaxy, with a corotation at 4.6~kpc, and an
inner Lindblad resonance (ILR) ring at radius 220~pc \citep{Fazeli2020}. The patchy ring could
correspond to a secondary bar/spiral, which has now been identified in the JWST images
\citep{Williams2024}.
 
The fact that the torus is very thin (axis ratio of 6.5,
and up to 10 taking into account the warped extension) does not favor the classical model of thick
dusty torus, as already questioned in several circumstances \citep[e.g.][]{Hoenig2019,Isbell2023}. It is
one of the first times when this geometrically thin character is clearly seen in any torus, 
except the H$_2$O masers thin disks around 0.5~pc in radius 
in other galaxies \citep[e.g.][]{Herrnstein1999, Gallimore2023}.

The velocity field looked deformed and slanted at low resolution,
because of the two components with different axes. In the model, mistaken by
including only one component, with the wrong orientation, 
 the predicted velocity dispersion was incorrect, and did not reproduce the
 observation.  The solution is not given either by the models from \citet{Bannikova2021},
 who propose a geometrically thick torus, inclined at i=60$^\circ$; the torus is made of
 optically thick clumps, and absorption produces the effect that the torus hotter and inner radii
 are only seen at some inclinations, at the origin of the velocity dispersion. However, molecular
 clumps on a given line of sight do not overlap in velocity and absorb each other,
 due to the velocity gradient, and molecular disks are
 considered globally optically thin \citep{Dickman1986}.

\subsection{Warping and misalignment}
 The molecular torus in NGC~1672 is clearly warped. Other tori might also be warped, but this is not easy to see because of their inclination. In NGC~613, the torus appears as an offset/lopsided molecular ring with two spirals joining at a position angle of PA=45$^\circ$. The spiral's SE junction coincides with a peak of emission.
 The oval shape looks somewhat truncated at SE, which might be due to bending. While
it is now well established that molecular tori are often kinematically decoupled
\citep{Garcia-Burillo2016,Combes2019,Tristram2022}, and have a dynamical state rather independent of their galaxy, they are still embedded in a potential well which is not yet completely spherical, as near the black hole, but has a restoring force toward the disk. It is then likely that they precess, and also warp, on a dynamical time-scale. They are still subject to their own self-gravity, since their masses are not negligible with respect to the super-massive black hole (SMBH).

It is possible to give an order of magnitude of the precessing rate as a function of the flattening of the potential. In a slightly oblate potential, the precession rate $\omega_p$ of a mass m in rotation with angular speed $\Omega$ at a radius r, in a plane tilted by an angle $\theta$ with the symmetry axis of the potential can be written as \citep{Sparke1996}:
\begin{equation}
\omega_p = \frac{1}{m r^2 \Omega} \frac{\partial V(r,\theta)}{\partial\, cos\theta}
\end{equation}
\noindent where $V$ is the potential energy of the mass m, averaged along its circular orbit. In a nearly spherical potential, with ellipticity $\eta$, assuming a nearly flat rotation curve, this can be approximated by: 
\begin{equation}
\omega_p  \propto\, \frac{\eta}{ r} cos\theta
\end{equation}
The derived precession period $ T_p = 2 \pi / \omega_p$ in such a slightly oblate potential is of the order of a few orbital periods, for $\theta < $ 40$^\circ$. The life-time of the warp will
depend on the differential precession as a function of radius, which is decreasing with radius,
unless $cos \theta \propto\, r$ to cancel the differential rates. Warps 
could then be as long-lived as the torus themselves in their outer parts. The life-time of the torus 
is limited by instabilities, that accelerate gas accretion toward the SMBH \citep[e.g.][]{Barker2014}.

Molecular tori are in the sphere of influence of their central black holes, with a nearly Keplerian potential, but also with some contribution of their self-gravity. In this region, given some pressure forces, there is a range of radii where the specific angular momentum is nearly constant, and where we could expect non-axisymmetric instabilities \citep{Papaloizou1984}, in particular since the PPI (Papaloizou-Pringle-Instability) persists if the specific angular momentum is not uniform \citep{Papaloizou1985}. Numerical simulations have shown the development of lopsided m=1 instabilities \citep{Barker2016, Donmez2017}, and in particular under some accretion perturbation \citep{Donmez2014}. In such conditions, where the instabilities can live several rotation periods, the gas is driven toward the center through the subsequent torques. The life-time of such a torus is of the order of a few Myr, and should be refilled through accretion from the rest of the disk.

Accretion is precisely one of the main mechanisms to produce misaligned nuclear disks.
When gas clouds are driven inward from the thicker large-scale disk, they arrive with randomized 
angular momentum \citep{Hopkins-align2012, Alcazar2021}.
Also gas can be acquired from a fountain effect, after star formation feedback, raining back into the
galaxy plane. Such phenomena have been found in N-body-hydro simulations, forming nuclear polar disks 
\cite[e.g.][]{Emsellem2015}.


\section{Conclusions}
\label{conclu}

We have presented high spatial resolution maps from ALMA cycle 7, of the two Seyfert galaxies
NGC~613 and NGC~1672. In Band 7, with the maximum baseline of 16~km, we have achieved a
beam of 15~mas in robust weighting for the CO(3-2) line, and 19~mas in natural weighting
for the HCO$^+$(4-3). The resolution corresponds to 1.2 and 0.8~pc on the NGC~613 
and NGC~1672 plane respectively. This is well below the sphere of influence
of their SMBH. Comparing the derived morphology and kinematics with
our previous low resolution maps obtained with more compact configurations,
we found several new features:
\begin{itemize}
\item 
the molecular tori that were barely resolved are now smaller; in NGC~613, it has been decomposed
in an actual torus, like a ring, with about the same inclination and position angle,
but is now 8~pc in radius. It appears as a lopsided ring, with a hole at 3~pc from the center, 
and with two nascent spiral
arms, joining the nuclear spiral structure. In NGC~1672, there are two components, the true molecular
torus, still kinematically decoupled from the galaxy disk, is now edge-on with another position
angle, and is strongly warped. It is embedded in a more face-on structure, with a different 
position angle.

\item 
the strong rotation around an axis completely different from the galaxy axis is confirmed in
both galaxies. Given the high resolution, we get velocities well inside the sphere
of influence of the black hole, where the other mass components of the galaxies,
essentially the bulge at pc-scale distance from the center, provide a negligible contribution.
Assuming circular orbits, it is possible to estimate directly a black hole mass M$_{BH}$.
The latter is perfectly consistent with previous estimates in both galaxies. The velocity 
gradients are now well resolved.  With high resolution, it is now realized that
 the torus in NGC~1672 is smaller, with a different inclination.

\item 
while NGC~613 shows an [OIII] ionized gas outflow in the direction of the radio jet, at 
the kpc-scale, and we see a molecular outflow in the same direction at 25~pc-scale,
we do not see the outflow at pc-scale. Instead we see line wings extending the rotation,
which is red-shifted in a direction opposite to the jet.

\item
the continuum emission in both galaxies is well resolved, and severely filtered out by the interferometer.
In NGC~613, its elongated shape may indicate the start of the radio jet.
The line emission is less filtered out, and we are only missing $\sim$ 10-40\% of
the flux.

\item 
The dense gas tracer HCO$^+$(4-3) line is well detected, and shows a high central concentration.
In the nucleus of NGC~613, the line is almost as strong as the CO(3-2) line. 
The CO(3-2)/HCO$^+$(4-3) ratio decreases rapidly with  decreasing radius, and reaches globally 2.4 in 
NGC~613 and 8 in the less concentrated nucleus NGC~1672. At low resolution, in the field of view
including the 300~pc ring, the ratio climbed up to 40 in NGC~613.

\item
 High resolution has enabled to fully resolve molecular tori, one of them is
 edge-on, geometrically thin, and warped. This does not favor the classical
 geometrically thick tori model. The circum-nuclear material, 
 that can obscure the AGN, is not a thick dusty torus, but a thin molecular torus. 
 It is able to obscure the AGN if edge-on, also aided by the polar dust 
 when present, which may widen the obscuring area. The thin torus may be short lived,
 by a few rotation periods, due to PPI instabilities, and the warp in its outer parts
 should have about the same life-time, due to differential precession in a
 nearly spherical potential. This short life-time ensures an efficient fueling
 of the SMBH, as soon as the gas has accumulated in the torus, through gravity torques.
 It is a natural explanation for the variability of the AGN, and may be of the
 changing look of some objects \citep[e.g.][]{Ricci2023}. Further fueling of the nucleus
 could then be weakened or stopped due to feedback, waiting for the next secular evolution cycle.
\end{itemize}

 These results show beautifully how much information ALMA and its high spatial
 resolution can provide to unveil the morphology and dynamics of the
 AGN circum-nuclear regions. Warping and misalignment are a clue to the transient nature
 of these tori, which may live a few dynamical time-scales. Mapping only at 
 lower angular resolution may give an incomplete impression
of the molecular gas kinematics and distribution close to galaxy nuclei. 


\begin{acknowledgements}
 We sincerely thank the referee for very useful and constructive comments, and a thorough reading of the manuscript.
The ALMA staff in Chile and ARC-people at IRAM are gratefully 
acknowledged for their help in the data reduction. 
SA gratefully acknowledges funding from the European Research Council (ERC)
under the European Union’s Horizon 2020 research and innovation programme (grant agreement
No. 789410).
AA acknowledges support from the European Union (WIDERA ExGal-Twin, GA 101158446)
and from the Agencia Estatal de Investigaci\'on of the Ministerio de Ciencia,
Innovaci\'on y Universidades (MCIU/AEI) under the grant ``Tracking active galactic nuclei
feedback from parsec to kiloparsec scales'', with reference PID2022$-$141105NB$-$I00 
and the European Regional Development Fund (ERDF).
VC acknowledges funding from the INAF Mini Grant 2022 program “Face-to-Face with the Local
Universe: ISM’s Empowerment (LOCAL).
SGB acknowledges support from the Spanish grant PID2022-138560NB-I00, funded by
MCIN/AEI/10.13039/501100011033/FEDER, EU.
LKH acknowledges funding from the INAF PRIN-SKA 2017 program 1.05.01.88.04. 
SV acknowledges support from the European Research Council (ERC) under the European Union’s Horizon
2020 research and innovation programme MOPPEX 833460.
This paper makes use of the following ALMA data: ADS/JAO.ALMA project 2019.0.00273.S
(PI: F. Combes). ALMA is a partnership of ESO (representing 
its member states), NSF (USA), and NINS (Japan), together with
NRC (Canada) and NSC and ASIAA (Taiwan), in cooperation with the Republic
 of Chile. The Joint ALMA Observatory is operated by ESO, AUI/NRAO,
and NAOJ. The National Radio Astronomy Observatory is a facility of the 
National Science Foundation operated under cooperative agreement by Associated 
Universities, Inc. We used observations made with the NASA/ESA Hubble 
Space Telescope, and obtained from the Hubble Legacy Archive, which is
a collaboration between the Space Telescope Science Institute (STScI/NASA),
the Space Telescope European Coordinating Facility (ST-ECF/ESA), and the
Canadian Astronomy Data center (CADC/NRC/CSA). This
work was supported by the Programme National Cosmology et Galaxies (PNCG)
of CNRS/INSU with INP and IN2P3, co- funded by CEA and CNES.
The data were processed using the Gildas package.
We made use of the NASA/IPAC Extragalactic Database (NED),
 and of the HyperLeda database (http://leda.univ-lyon1.fr).
\end{acknowledgements}

\bibliographystyle{aa}
\bibliography{NUGA.bib}

\begin{thebibliography}{90}
\expandafter\ifx\csname natexlab\endcsname\relax\def\natexlab#1{#1}\fi

\bibitem[{{Alonso-Herrero} {et~al.}(2019){Alonso-Herrero},
  {Garc{\'\i}a-Burillo}, {Pereira-Santaella}, {Davies}, {Combes},
  {Vestergaard}, {Raimundo}, {Bunker}, {D{\'\i}az-Santos}, {Gandhi},
  {Garc{\'\i}a-Bernete}, {Hicks}, {H{\"o}nig}, {Hunt}, {Imanishi}, {Izumi},
  {Levenson}, {Maciejewski}, {Packham}, {Ramos Almeida}, {Ricci}, {Rigopoulou},
  {Roche}, {Rosario}, {Schartmann}, {Usero}, \& {Ward}}]{Alonso-Herrero2019}
{Alonso-Herrero}, A., {Garc{\'\i}a-Burillo}, S., {Pereira-Santaella}, M.,
  {et~al.} 2019, \aap, 628, A65

\bibitem[{{Alonso Herrero} {et~al.}(2023){Alonso Herrero},
  {Garc{\'\i}a-Burillo}, {Pereira-Santaella}, {Shimizu}, {Combes}, {Hicks},
  {Davies}, {Ramos Almeida}, {Garc{\'\i}a-Bernete}, {H{\"o}nig}, {Levenson},
  {Packham}, {Bellocchi}, {Hunt}, {Imanishi}, {Ricci}, \&
  {Roche}}]{Alonso-Herrero2023}
{Alonso Herrero}, A., {Garc{\'\i}a-Burillo}, S., {Pereira-Santaella}, M.,
  {et~al.} 2023, \aap, 675, A88

\bibitem[{{Angl{\'e}s-Alc{\'a}zar} {et~al.}(2021){Angl{\'e}s-Alc{\'a}zar},
  {Quataert}, {Hopkins}, {Somerville}, {Hayward}, {Faucher-Gigu{\`e}re},
  {Bryan}, {Kere{\v{s}}}, {Hernquist}, \& {Stone}}]{Alcazar2021}
{Angl{\'e}s-Alc{\'a}zar}, D., {Quataert}, E., {Hopkins}, P.~F., {et~al.} 2021,
  \apj, 917, 53

\bibitem[{{Antonucci} \& {Miller}(1985)}]{Antonucci1985}
{Antonucci}, R.~R.~J. \& {Miller}, J.~S. 1985, \apj, 297, 621

\bibitem[{{Asmus}(2019)}]{Asmus2019}
{Asmus}, D. 2019, \mnras, 489, 2177

\bibitem[{{Asmus} {et~al.}(2016){Asmus}, {H{\"o}nig}, \& {Gandhi}}]{Asmus2016}
{Asmus}, D., {H{\"o}nig}, S.~F., \& {Gandhi}, P. 2016, \apj, 822, 109

\bibitem[{{Audibert} {et~al.}(2019){Audibert}, {Combes}, {Garc{\'\i}a-Burillo},
  {Hunt}, {Eckart}, {Aalto}, {Casasola}, {Boone}, {Krips}, {Viti}, {Muller},
  {Dasyra}, {van der Werf}, \& {Mart{\'\i}n}}]{Audibert2019}
{Audibert}, A., {Combes}, F., {Garc{\'\i}a-Burillo}, S., {et~al.} 2019, \aap,
  632, A33

\bibitem[{{Audibert} {et~al.}(2021){Audibert}, {Combes}, {Garc{\'\i}a-Burillo},
  {Hunt}, {Eckart}, {Aalto}, {Casasola}, {Boone}, {Krips}, {Viti}, {Muller},
  {Dasyra}, {van der Werf}, \& {Mart{\'\i}n}}]{Audibert2021}
{Audibert}, A., {Combes}, F., {Garc{\'\i}a-Burillo}, S., {et~al.} 2021, \aap,
  656, A60

\bibitem[{{Bannikova} {et~al.}(2021){Bannikova}, {Sergeyev}, {Akerman},
  {Berczik}, {Ishchenko}, {Capaccioli}, \& {Akhmetov}}]{Bannikova2021}
{Bannikova}, E.~Y., {Sergeyev}, A.~V., {Akerman}, N.~A., {et~al.} 2021, \mnras,
  503, 1459

\bibitem[{{Barker} \& {Ogilvie}(2014)}]{Barker2014}
{Barker}, A.~J. \& {Ogilvie}, G.~I. 2014, \mnras, 445, 2637

\bibitem[{{Barker} \& {Ogilvie}(2016)}]{Barker2016}
{Barker}, A.~J. \& {Ogilvie}, G.~I. 2016, \mnras, 458, 3739

\bibitem[{{Bolatto} {et~al.}(2013){Bolatto}, {Wolfire}, \&
  {Leroy}}]{Bolatto2013}
{Bolatto}, A.~D., {Wolfire}, M., \& {Leroy}, A.~K. 2013, \araa, 51, 207

\bibitem[{{Casasola} {et~al.}(2011){Casasola}, {Hunt}, {Combes},
  {Garc{\'{\i}}a-Burillo}, \& {Neri}}]{Casasola2011}
{Casasola}, V., {Hunt}, L.~K., {Combes}, F., {Garc{\'{\i}}a-Burillo}, S., \&
  {Neri}, R. 2011, \aap, 527, A92

\bibitem[{{Castangia} {et~al.}(2013){Castangia}, {Panessa}, {Henkel}, {Kadler},
  \& {Tarchi}}]{Castangia2013}
{Castangia}, P., {Panessa}, F., {Henkel}, C., {Kadler}, M., \& {Tarchi}, A.
  2013, \mnras, 436, 3388

\bibitem[{{Castangia} {et~al.}(2008){Castangia}, {Tarchi}, {Henkel}, \&
  {Menten}}]{Castangia2008}
{Castangia}, P., {Tarchi}, A., {Henkel}, C., \& {Menten}, K.~M. 2008, \aap,
  479, 111

\bibitem[{{Cicone} {et~al.}(2014){Cicone}, {Maiolino}, {Sturm},
  {Graci{\'a}-Carpio}, {Feruglio}, {Neri}, {Aalto}, {Davies}, {Fiore},
  {Fischer}, {Garc{\'{\i}}a-Burillo}, {Gonz{\'a}lez-Alfonso},
  {Hailey-Dunsheath}, {Piconcelli}, \& {Veilleux}}]{Cicone2014}
{Cicone}, C., {Maiolino}, R., {Sturm}, E., {et~al.} 2014, \aap, 562, A21

\bibitem[{{Combes}(2012)}]{Combes2012}
{Combes}, F. 2012, in Journal of Physics Conference Series, Vol. 372, Journal
  of Physics Conference Series (IOP), 012041

\bibitem[{{Combes} {et~al.}(2019){Combes}, {Garc{\'\i}a-Burillo}, {Audibert},
  {Hunt}, {Eckart}, {Aalto}, {Casasola}, {Boone}, {Krips}, {Viti}, {Sakamoto},
  {Muller}, {Dasyra}, {van der Werf}, \& {Martin}}]{Combes2019}
{Combes}, F., {Garc{\'\i}a-Burillo}, S., {Audibert}, A., {et~al.} 2019, \aap,
  623, A79

\bibitem[{{Combes} {et~al.}(2013){Combes}, {Garc{\'{\i}}a-Burillo}, {Casasola},
  {Hunt}, {Krips}, {Baker}, {Boone}, {Eckart}, {Marquez}, {Neri}, {Schinnerer},
  \& {Tacconi}}]{Combes2013}
{Combes}, F., {Garc{\'{\i}}a-Burillo}, S., {Casasola}, V., {et~al.} 2013, \aap,
  558, A124

\bibitem[{{Combes} {et~al.}(2014){Combes}, {Garc{\'{\i}}a-Burillo}, {Casasola},
  {Hunt}, {Krips}, {Baker}, {Boone}, {Eckart}, {Marquez}, {Neri}, {Schinnerer},
  \& {Tacconi}}]{Combes2014}
{Combes}, F., {Garc{\'{\i}}a-Burillo}, S., {Casasola}, V., {et~al.} 2014, \aap,
  565, A97

\bibitem[{{de Naray} {et~al.}(2000){de Naray}, {Brandt}, {Halpern}, \&
  {Iwasawa}}]{deNaray2000}
{de Naray}, P.~J., {Brandt}, W.~N., {Halpern}, J.~P., \& {Iwasawa}, K. 2000,
  \aj, 119, 612

\bibitem[{{D{\'{\i}}az} {et~al.}(1999){D{\'{\i}}az}, {Carranza}, {Dottori}, \&
  {Goldes}}]{Diaz1999}
{D{\'{\i}}az}, R., {Carranza}, G., {Dottori}, H., \& {Goldes}, G. 1999, \apj,
  512, 623

\bibitem[{{Dickman} {et~al.}(1986){Dickman}, {Snell}, \&
  {Schloerb}}]{Dickman1986}
{Dickman}, R.~L., {Snell}, R.~L., \& {Schloerb}, F.~P. 1986, \apj, 309, 326

\bibitem[{{D{\"o}nmez}(2014)}]{Donmez2014}
{D{\"o}nmez}, O. 2014, \mnras, 438, 846

\bibitem[{{Donmez}(2017)}]{Donmez2017}
{Donmez}, O. 2017, Modern Physics Letters A, 32, 1750108

\bibitem[{{Dubois} {et~al.}(2013){Dubois}, {Gavazzi}, {Peirani}, \&
  {Silk}}]{Dubois2013}
{Dubois}, Y., {Gavazzi}, R., {Peirani}, S., \& {Silk}, J. 2013, \mnras, 433,
  3297

\bibitem[{{Emsellem} {et~al.}(2015){Emsellem}, {Renaud}, {Bournaud},
  {Elmegreen}, {Combes}, \& {Gabor}}]{Emsellem2015}
{Emsellem}, E., {Renaud}, F., {Bournaud}, F., {et~al.} 2015, \mnras, 446, 2468

\bibitem[{{Esposito} {et~al.}(2024){Esposito}, {Alonso-Herrero},
  {Garc{\'\i}a-Burillo}, {Casasola}, {Combes}, {Dallacasa}, {Davies},
  {Garc{\'\i}a-Bernete}, {Garc{\'\i}a-Lorenzo}, {Hermosa Mu{\~n}oz}, {de
  Arriba}, {Pereira-Santaella}, {Pozzi}, {Ramos Almeida}, {Shimizu}, {Vallini},
  {Bellocchi}, {Gonz{\'a}lez-Mart{\'\i}n}, {Hicks}, {H{\"o}nig}, {Labiano},
  {Levenson}, {Ricci}, \& {Rosario}}]{Esposito2024}
{Esposito}, F., {Alonso-Herrero}, A., {Garc{\'\i}a-Burillo}, S., {et~al.} 2024,
  \aap, 686, A46

\bibitem[{{Fazeli} {et~al.}(2020){Fazeli}, {Eckart}, {Busch}, {Yttergren},
  {Combes}, {Misquitta}, \& {Straubmeier}}]{Fazeli2020}
{Fazeli}, N., {Eckart}, A., {Busch}, G., {et~al.} 2020, \aap, 638, A36

\bibitem[{{Fiore} {et~al.}(2017){Fiore}, {Feruglio}, {Shankar}, {Bischetti},
  {Bongiorno}, {Brusa}, {Carniani}, {Cicone}, {Duras}, {Lamastra}, {Mainieri},
  {Marconi}, {Menci}, {Maiolino}, {Piconcelli}, {Vietri}, \&
  {Zappacosta}}]{Fiore2017}
{Fiore}, F., {Feruglio}, C., {Shankar}, F., {et~al.} 2017, \aap, 601, A143

\bibitem[{{Fluetsch} {et~al.}(2019){Fluetsch}, {Maiolino}, {Carniani},
  {Marconi}, {Cicone}, {Bourne}, {Costa}, {Fabian}, {Ishibashi}, \&
  {Venturi}}]{Fluetsch2019}
{Fluetsch}, A., {Maiolino}, R., {Carniani}, S., {et~al.} 2019, \mnras, 483,
  4586

\bibitem[{{Gabor} \& {Bournaud}(2014)}]{Gabor2014}
{Gabor}, J.~M. \& {Bournaud}, F. 2014, \mnras, 441, 1615

\bibitem[{{Gallimore} \& {Impellizzeri}(2023)}]{Gallimore2023}
{Gallimore}, J.~F. \& {Impellizzeri}, C.~M.~V. 2023, \apj, 951, 109

\bibitem[{{G{\'a}mez Rosas} {et~al.}(2025){G{\'a}mez Rosas}, {van der Werf},
  {Gallimore}, {Impellizzeri}, {Jaffe}, {Garc{\'\i}a-Burillo}, {Aalto},
  {Burtscher}, {Casasola}, {Combes}, {Henkel}, {M{\'a}rquez}, {Mart{\'\i}n},
  {Ramos Almeida}, {Viti}, \& {Fuente}}]{Gamez-Rosas2025}
{G{\'a}mez Rosas}, V., {van der Werf}, P., {Gallimore}, J.~F., {et~al.} 2025,
  \aap, 699, A187

\bibitem[{{Garc{\'\i}a-Bernete} {et~al.}(2021){Garc{\'\i}a-Bernete},
  {Alonso-Herrero}, {Garc{\'\i}a-Burillo}, {Pereira-Santaella},
  {Garc{\'\i}a-Lorenzo}, {Carrera}, {Rigopoulou}, {Ramos Almeida}, {Villar
  Mart{\'\i}n}, {Gonz{\'a}lez-Mart{\'\i}n}, {Hicks}, {Labiano}, {Ricci}, \&
  {Mateos}}]{Garcia-Bernete2021}
{Garc{\'\i}a-Bernete}, I., {Alonso-Herrero}, A., {Garc{\'\i}a-Burillo}, S.,
  {et~al.} 2021, \aap, 645, A21

\bibitem[{{Garc{\'\i}a-Burillo} {et~al.}(2021){Garc{\'\i}a-Burillo},
  {Alonso-Herrero}, {Ramos Almeida}, {Gonz{\'a}lez-Mart{\'\i}n}, {Combes},
  {Usero}, {H{\"o}nig}, {Querejeta}, {Hicks}, {Hunt}, {Rosario}, {Davies},
  {Boorman}, {Bunker}, {Burtscher}, {Colina}, {D{\'\i}az-Santos}, {Gandhi},
  {Garc{\'\i}a-Bernete}, {Garc{\'\i}a-Lorenzo}, {Ichikawa}, {Imanishi},
  {Izumi}, {Labiano}, {Levenson}, {L{\'o}pez-Rodr{\'\i}guez}, {Packham},
  {Pereira-Santaella}, {Ricci}, {Rigopoulou}, {Rouan}, {Shimizu}, {Stalevski},
  {Wada}, \& {Williamson}}]{Garcia-Burillo2021}
{Garc{\'\i}a-Burillo}, S., {Alonso-Herrero}, A., {Ramos Almeida}, C., {et~al.}
  2021, \aap, 652, A98

\bibitem[{{Garc{\'\i}a-Burillo} {et~al.}(2003){Garc{\'\i}a-Burillo}, {Combes},
  {Hunt}, {Boone}, {Baker}, {Tacconi}, {Eckart}, {Neri}, {Leon}, {Schinnerer},
  \& {Englmaier}}]{Garcia-Burillo2003}
{Garc{\'\i}a-Burillo}, S., {Combes}, F., {Hunt}, L.~K., {et~al.} 2003, \aap,
  407, 485

\bibitem[{{Garc{\'\i}a-Burillo} {et~al.}(2019){Garc{\'\i}a-Burillo}, {Combes},
  {Ramos Almeida}, {Usero}, {Alonso-Herrero}, {Hunt}, {Rouan}, {Aalto},
  {Querejeta}, {Viti}, {van der Werf}, {Vives-Arias}, {Fuente}, {Colina},
  {Mart{\'\i}n-Pintado}, {Henkel}, {Mart{\'\i}n}, {Krips}, {Gratadour}, {Neri},
  \& {Tacconi}}]{Garcia-Burillo2019}
{Garc{\'\i}a-Burillo}, S., {Combes}, F., {Ramos Almeida}, C., {et~al.} 2019,
  \aap, 632, A61

\bibitem[{{Garc{\'{\i}}a-Burillo} {et~al.}(2016){Garc{\'{\i}}a-Burillo},
  {Combes}, {Ramos Almeida}, {Usero}, {Krips}, {Alonso-Herrero}, {Aalto},
  {Casasola}, {Hunt}, {Mart{\'{\i}}n}, {Viti}, {Colina}, {Costagliola},
  {Eckart}, {Fuente}, {Henkel}, {M{\'a}rquez}, {Neri}, {Schinnerer}, {Tacconi},
  \& {van der Werf}}]{Garcia-Burillo2016}
{Garc{\'{\i}}a-Burillo}, S., {Combes}, F., {Ramos Almeida}, C., {et~al.} 2016,
  \apjl, 823, L12

\bibitem[{{Garc{\'{\i}}a-Burillo} {et~al.}(2005){Garc{\'{\i}}a-Burillo},
  {Combes}, {Schinnerer}, {Boone}, \& {Hunt}}]{Garcia-Burillo2005}
{Garc{\'{\i}}a-Burillo}, S., {Combes}, F., {Schinnerer}, E., {Boone}, F., \&
  {Hunt}, L.~K. 2005, \aap, 441, 1011

\bibitem[{{Garc{\'{\i}}a-Burillo} {et~al.}(2014){Garc{\'{\i}}a-Burillo},
  {Combes}, {Usero}, {Aalto}, {Krips}, {Viti}, {Alonso-Herrero}, {Hunt},
  {Schinnerer}, {Baker}, {Boone}, {Casasola}, {Colina}, {Costagliola},
  {Eckart}, {Fuente}, {Henkel}, {Labiano}, {Mart{\'{\i}}n}, {M{\'a}rquez},
  {Muller}, {Planesas}, {Ramos Almeida}, {Spaans}, {Tacconi}, \& {van der
  Werf}}]{Garcia-Burillo2014}
{Garc{\'{\i}}a-Burillo}, S., {Combes}, F., {Usero}, A., {et~al.} 2014, \aap,
  567, A125

\bibitem[{{Garc{\'\i}a-Burillo} {et~al.}(2024){Garc{\'\i}a-Burillo}, {Hicks},
  {Alonso-Herrero}, {Pereira-Santaella}, {Usero}, {Querejeta},
  {Gonz{\'a}lez-Mart{\'\i}n}, {Delaney}, {Ramos Almeida}, {Combes},
  {Angl{\'e}s-Alc{\'a}zar}, {Audibert}, {Bellocchi}, {Davies}, {Davis},
  {Elford}, {Garc{\'\i}a-Bernete}, {H{\"o}nig}, {Labiano}, {Leist}, {Levenson},
  {L{\'o}pez-Rodr{\'\i}guez}, {Mercedes-Feliz}, {Packham}, {Ricci}, {Rosario},
  {Shimizu}, {Stalevski}, \& {Zhang}}]{Garcia-Burillo2024}
{Garc{\'\i}a-Burillo}, S., {Hicks}, E.~K.~S., {Alonso-Herrero}, A., {et~al.}
  2024, \aap, 689, A347

\bibitem[{{Goesaert} {et~al.}(2025){Goesaert}, {Tristram}, {Impellizzeri},
  {Hygate}, {Venselaar}, {Wang}, \& {Zhang}}]{Goesaert2025}
{Goesaert}, W.~M., {Tristram}, K. R.~W., {Impellizzeri}, C.~M.~V., {et~al.}
  2025, arXiv e-prints, arXiv:2510.05199

\bibitem[{{Gorski} {et~al.}(2024){Gorski}, {Aalto}, {K{\"o}nig}, {Wethers},
  {Yang}, {Muller}, {Onishi}, {Sato}, {Falstad}, {Mangum}, {Linden}, {Combes},
  {Mart{\'\i}n}, {Imanishi}, {Wada}, {Barcos-Mu{\~n}oz}, {Stanley},
  {Garc{\'\i}a-Burillo}, {van der Werf}, {Evans}, {Henkel}, {Viti}, {Harada},
  {D{\'\i}az-Santos}, {Gallagher}, \& {Gonz{\'a}lez-Alfonso}}]{Gorski2024}
{Gorski}, M.~D., {Aalto}, S., {K{\"o}nig}, S., {et~al.} 2024, \aap, 684, L11

\bibitem[{{Gratadour} {et~al.}(2015){Gratadour}, {Rouan}, {Grosset},
  {Boccaletti}, \& {Cl{\'e}net}}]{Gratadour2015}
{Gratadour}, D., {Rouan}, D., {Grosset}, L., {Boccaletti}, A., \& {Cl{\'e}net},
  Y. 2015, \aap, 581, L8

\bibitem[{{GRAVITY Collaboration} {et~al.}(2020){GRAVITY Collaboration},
  {Pfuhl}, {Davies}, {Dexter}, {Netzer}, {H{\"o}nig}, {Lutz}, {Schartmann},
  {Sturm}, {Amorim}, {Brandner}, {Cl{\'e}net}, {de Zeeuw}, {Eckart},
  {Eisenhauer}, {F{\"o}rster Schreiber}, {Gao}, {Garcia}, {Genzel},
  {Gillessen}, {Gratadour}, {Kishimoto}, {Lacour}, {Millour}, {Ott}, {Paumard},
  {Perraut}, {Perrin}, {Peterson}, {Petrucci}, {Prieto}, {Rouan}, {Shangguan},
  {Shimizu}, {Sternberg}, {Straub}, {Straubmeier}, {Tacconi}, {Tristram},
  {Vermot}, {Waisberg}, {Widmann}, \& {Woillez}}]{GRAVITY2020}
{GRAVITY Collaboration}, {Pfuhl}, O., {Davies}, R., {et~al.} 2020, \aap, 634,
  A1

\bibitem[{{Greve} {et~al.}(2009){Greve}, {Papadopoulos}, {Gao}, \&
  {Radford}}]{Greve2009}
{Greve}, T.~R., {Papadopoulos}, P.~P., {Gao}, Y., \& {Radford}, S.~J.~E. 2009,
  \apj, 692, 1432

\bibitem[{{Guilloteau} \& {Lucas}(2000)}]{Guilloteau2000}
{Guilloteau}, S. \& {Lucas}, R. 2000, in Astronomical Society of the Pacific
  Conference Series, Vol. 217, Imaging at Radio through Submillimeter
  Wavelengths, ed. J.~G. {Mangum} \& S.~J.~E. {Radford}, 299

\bibitem[{{Haidar} {et~al.}(2024){Haidar}, {Rosario}, {Alonso-Herrero},
  {Pereira-Santaella}, {Garc{\'\i}a-Bernete}, {Campbell}, {H{\"o}nig}, {Ramos
  Almeida}, {Hicks}, {Delaney}, {Davies}, {Ricci}, {Harrison}, {Leist},
  {Lopez-Rodriguez}, {Garcia-Burillo}, {Zhang}, {Packham}, {Gandhi},
  {Audibert}, {Bellocchi}, {Boorman}, {Bunker}, {Combes}, {Diaz Santos},
  {Donnan}, {Gonzalez Martin}, {Hermosa Mu{\~n}oz}, {Charidis}, {Labiano},
  {Levenson}, {May}, {Rigopoulou}, {Rodriguez Ardila}, {Shimizu}, {Stalevski},
  \& {Ward}}]{Haidar2024}
{Haidar}, H., {Rosario}, D.~J., {Alonso-Herrero}, A., {et~al.} 2024, \mnras,
  532, 4645

\bibitem[{{Herrnstein} {et~al.}(1999){Herrnstein}, {Moran}, {Greenhill},
  {Diamond}, {Inoue}, {Nakai}, {Miyoshi}, {Henkel}, \&
  {Riess}}]{Herrnstein1999}
{Herrnstein}, J.~R., {Moran}, J.~M., {Greenhill}, L.~J., {et~al.} 1999, \nat,
  400, 539

\bibitem[{{H{\"o}nig}(2019)}]{Hoenig2019}
{H{\"o}nig}, S.~F. 2019, \apj, 884, 171

\bibitem[{{H{\"o}nig} \& {Kishimoto}(2010)}]{Hoenig2010}
{H{\"o}nig}, S.~F. \& {Kishimoto}, M. 2010, \aap, 523, A27

\bibitem[{{Hopkins} {et~al.}(2024){Hopkins}, {Grudic}, {Su}, {Wellons},
  {Angles-Alcazar}, {Steinwandel}, {Guszejnov}, {Murray}, {Faucher-Giguere},
  {Quataert}, \& {Keres}}]{Hopkins2024}
{Hopkins}, P.~F., {Grudic}, M.~Y., {Su}, K.-Y., {et~al.} 2024, The Open Journal
  of Astrophysics, 7, 18

\bibitem[{{Hopkins} {et~al.}(2012){Hopkins}, {Hernquist}, {Hayward}, \&
  {Narayanan}}]{Hopkins-align2012}
{Hopkins}, P.~F., {Hernquist}, L., {Hayward}, C.~C., \& {Narayanan}, D. 2012,
  \mnras, 425, 1121

\bibitem[{{Hopkins} \& {Quataert}(2010)}]{Hopkins2010}
{Hopkins}, P.~F. \& {Quataert}, E. 2010, \mnras, 407, 1529

\bibitem[{{Hummel} \& {Jorsater}(1992)}]{Hummel1992}
{Hummel}, E. \& {Jorsater}, S. 1992, \aap, 261, 85

\bibitem[{{Hummel} {et~al.}(1987){Hummel}, {Jorsater}, {Lindblad}, \&
  {Sandqvist}}]{Hummel1987}
{Hummel}, E., {Jorsater}, S., {Lindblad}, P.~O., \& {Sandqvist}, A. 1987, \aap,
  172, 51

\bibitem[{{Imanishi} {et~al.}(2016){Imanishi}, {Nakanishi}, \&
  {Izumi}}]{Imanishi2016}
{Imanishi}, M., {Nakanishi}, K., \& {Izumi}, T. 2016, \apjl, 822, L10

\bibitem[{{Imanishi} {et~al.}(2020){Imanishi}, {Nguyen}, {Wada}, {Hagiwara},
  {Iguchi}, {Izumi}, {Kawakatu}, {Nakanishi}, \& {Onishi}}]{Imanishi2020}
{Imanishi}, M., {Nguyen}, D.~D., {Wada}, K., {et~al.} 2020, \apj, 902, 99

\bibitem[{{Impellizzeri} {et~al.}(2019){Impellizzeri}, {Gallimore}, {Baum},
  {Elitzur}, {Davies}, {Lutz}, {Maiolino}, {Marconi}, {Nikutta}, {O'Dea}, \&
  {Sani}}]{Impellizzeri2019}
{Impellizzeri}, C.~M.~V., {Gallimore}, J.~F., {Baum}, S.~A., {et~al.} 2019,
  \apjl, 884, L28

\bibitem[{{Isbell} {et~al.}(2023){Isbell}, {Pott}, {Meisenheimer}, {Stalevski},
  {Tristram}, {Leftley}, {Asmus}, {Weigelt}, {G{\'a}mez Rosas}, {Petrov},
  {Jaffe}, {Hofmann}, {Henning}, \& {Lopez}}]{Isbell2023}
{Isbell}, J.~W., {Pott}, J.~U., {Meisenheimer}, K., {et~al.} 2023, \aap, 678,
  A136

\bibitem[{{Izumi} {et~al.}(2016){Izumi}, {Kawakatu}, \& {Kohno}}]{Izumi2016}
{Izumi}, T., {Kawakatu}, N., \& {Kohno}, K. 2016, \apj, 827, 81

\bibitem[{{Jenkins} {et~al.}(2011){Jenkins}, {Brandt}, {Colbert}, {Koribalski},
  {Kuntz}, {Levan}, {Ojha}, {Roberts}, {Ward}, \& {Zezas}}]{Jenkins2011}
{Jenkins}, L.~P., {Brandt}, W.~N., {Colbert}, E.~J.~M., {et~al.} 2011, \apj,
  734, 33

\bibitem[{{Jungwiert} {et~al.}(1997){Jungwiert}, {Combes}, \&
  {Axon}}]{Jungwiert1997}
{Jungwiert}, B., {Combes}, F., \& {Axon}, D.~J. 1997, \aaps, 125, 479

\bibitem[{{Kameno} {et~al.}(2020){Kameno}, {Sawada-Satoh}, {Impellizzeri},
  {Espada}, {Nakai}, {Sugai}, {Terashima}, {Kohno}, {Lee}, \&
  {Mart{\'\i}n}}]{Kameno2020}
{Kameno}, S., {Sawada-Satoh}, S., {Impellizzeri}, C.~M.~V., {et~al.} 2020,
  \apj, 895, 73

\bibitem[{{Kewley} {et~al.}(2000){Kewley}, {Heisler}, {Dopita}, {Sutherland},
  {Norris}, {Reynolds}, \& {Lumsden}}]{Kewley2000}
{Kewley}, L.~J., {Heisler}, C.~A., {Dopita}, M.~A., {et~al.} 2000, \apj, 530,
  704

\bibitem[{{Kondratko} {et~al.}(2006){Kondratko}, {Greenhill}, {Moran},
  {Lovell}, {Kuiper}, {Jauncey}, {Cameron}, {G{\'o}mez},
  {Garc{\'{\i}}a-Mir{\'o}}, {Moll}, {de Gregorio-Monsalvo}, \&
  {Jim{\'e}nez-Bail{\'o}n}}]{Kondratko2006}
{Kondratko}, P.~T., {Greenhill}, L.~J., {Moran}, J.~M., {et~al.} 2006, \apj,
  638, 100

\bibitem[{{Koudmani} {et~al.}(2022){Koudmani}, {Sijacki}, \&
  {Smith}}]{Koudmani2022}
{Koudmani}, S., {Sijacki}, D., \& {Smith}, M.~C. 2022, \mnras, 516, 2112

\bibitem[{{Leftley} {et~al.}(2019){Leftley}, {H{\"o}nig}, {Asmus}, {Tristram},
  {Gandhi}, {Kishimoto}, {Venanzi}, \& {Williamson}}]{Leftley2019}
{Leftley}, J.~H., {H{\"o}nig}, S.~F., {Asmus}, D., {et~al.} 2019, \apj, 886, 55

\bibitem[{{Lutz} {et~al.}(2020){Lutz}, {Sturm}, {Janssen}, {Veilleux}, {Aalto},
  {Cicone}, {Contursi}, {Davies}, {Feruglio}, {Fischer}, {Fluetsch},
  {Garcia-Burillo}, {Genzel}, {Gonz{\'a}lez-Alfonso}, {Graci{\'a}-Carpio},
  {Herrera-Camus}, {Maiolino}, {Schruba}, {Shimizu}, {Sternberg}, {Tacconi}, \&
  {Wei{\ss}}}]{Lutz2020}
{Lutz}, D., {Sturm}, E., {Janssen}, A., {et~al.} 2020, \aap, 633, A134

\bibitem[{{Mahieu} {et~al.}(2012){Mahieu}, {Maier}, {Lazareff}, {Navarrini},
  {Celestin}, {Chalain}, {Geoffroy}, {Laslaz}, \& {Perrin}}]{Mahieu2012}
{Mahieu}, S., {Maier}, D., {Lazareff}, B., {et~al.} 2012, IEEE Transactions on
  Terahertz Science and Technology, 2, 29

\bibitem[{{Malkan} {et~al.}(1998){Malkan}, {Gorjian}, \& {Tam}}]{Malkan1998}
{Malkan}, M.~A., {Gorjian}, V., \& {Tam}, R. 1998, \apjs, 117, 25

\bibitem[{{McMullin} {et~al.}(2007){McMullin}, {Waters}, {Schiebel}, {Young},
  \& {Golap}}]{McMullin2007}
{McMullin}, J.~P., {Waters}, B., {Schiebel}, D., {Young}, W., \& {Golap}, K.
  2007, in Astronomical Society of the Pacific Conference Series, Vol. 376,
  Astronomical Data Analysis Software and Systems XVI, ed. R.~A. {Shaw},
  F.~{Hill}, \& D.~J. {Bell}, 127

\bibitem[{{Merritt}(2004)}]{Merritt2004}
{Merritt}, D. 2004, Coevolution of Black Holes and Galaxies, 263

\bibitem[{{Miyamoto} {et~al.}(2017){Miyamoto}, {Nakai}, {Seta}, {Salak},
  {Nagai}, \& {Kaneko}}]{Miyamoto2017}
{Miyamoto}, Y., {Nakai}, N., {Seta}, M., {et~al.} 2017, \pasj, 69, 83

\bibitem[{{Nenkova} {et~al.}(2008){Nenkova}, {Sirocky}, {Nikutta},
  {Ivezi{\'c}}, \& {Elitzur}}]{Nenkova2008}
{Nenkova}, M., {Sirocky}, M.~M., {Nikutta}, R., {Ivezi{\'c}}, {\v{Z}}., \&
  {Elitzur}, M. 2008, \apj, 685, 160

\bibitem[{{Papaloizou} \& {Pringle}(1984)}]{Papaloizou1984}
{Papaloizou}, J.~C.~B. \& {Pringle}, J.~E. 1984, \mnras, 208, 721

\bibitem[{{Papaloizou} \& {Pringle}(1985)}]{Papaloizou1985}
{Papaloizou}, J.~C.~B. \& {Pringle}, J.~E. 1985, \mnras, 213, 799

\bibitem[{{Ramos Almeida} \& {Ricci}(2017)}]{Ramos2017}
{Ramos Almeida}, C. \& {Ricci}, C. 2017, Nature Astronomy, 1, 679

\bibitem[{{Ricci} \& {Trakhtenbrot}(2023)}]{Ricci2023}
{Ricci}, C. \& {Trakhtenbrot}, B. 2023, Nature Astronomy, 7, 1282

\bibitem[{{Silva-Lima} {et~al.}(2025){Silva-Lima}, {Gadotti}, {Martins},
  {Kolcu}, {Coelho}, {Fragkoudi}, {Kim}, {de S{\'a}-Freitas},
  {Falc{\'o}n-Barroso}, {de Lorenzo-C{\'a}ceres}, {M{\'e}ndez-Abreu},
  {Neumann}, {Querejeta}, \& {S{\'a}nchez-Bl{\'a}zquez}}]{Silva-Lima2025}
{Silva-Lima}, L.~A., {Gadotti}, D.~A., {Martins}, L.~P., {et~al.} 2025, \mnras,
  540, 2787

\bibitem[{{Sparke}(1996)}]{Sparke1996}
{Sparke}, L.~S. 1996, \apj, 473, 810

\bibitem[{{Steer} {et~al.}(2017){Steer}, {Madore}, {Mazzarella}, {Schmitz},
  {Corwin}, {Chan}, {Ebert}, {Helou}, {Baker}, {Chen}, {Frayer}, {Jacobson},
  {Lo}, {Ogle}, {Pevunova}, \& {Terek}}]{Steer2017}
{Steer}, I., {Madore}, B.~F., {Mazzarella}, J.~M., {et~al.} 2017, \aj, 153, 37

\bibitem[{{Tacconi} {et~al.}(2018){Tacconi}, {Genzel}, {Saintonge}, {Combes},
  {Garc{\'\i}a-Burillo}, {Neri}, {Bolatto}, {Contini}, {F{\"o}rster Schreiber},
  {Lilly}, {Lutz}, {Wuyts}, {Accurso}, {Boissier}, {Boone}, {Bouch{\'e}},
  {Bournaud}, {Burkert}, {Carollo}, {Cooper}, {Cox}, {Feruglio}, {Freundlich},
  {Herrera-Camus}, {Juneau}, {Lippa}, {Naab}, {Renzini}, {Salome}, {Sternberg},
  {Tadaki}, {{\"U}bler}, {Walter}, {Weiner}, \& {Weiss}}]{Tacconi2018}
{Tacconi}, L.~J., {Genzel}, R., {Saintonge}, A., {et~al.} 2018, \apj, 853, 179

\bibitem[{{Tristram} {et~al.}(2022){Tristram}, {Impellizzeri}, {Zhang},
  {Villard}, {Henkel}, {Viti}, {Burtscher}, {Combes}, {Garc{\'\i}a-Burillo},
  {Mart{\'\i}n}, {Meisenheimer}, \& {van der Werf}}]{Tristram2022}
{Tristram}, K. R.~W., {Impellizzeri}, C.~M.~V., {Zhang}, Z.-Y., {et~al.} 2022,
  \aap, 664, A142

\bibitem[{{Urry} \& {Padovani}(1995)}]{Urry1995}
{Urry}, C.~M. \& {Padovani}, P. 1995, \pasp, 107, 803

\bibitem[{{Usero} {et~al.}(2015){Usero}, {Leroy}, {Walter}, {Schruba},
  {Garc{\'\i}a-Burillo}, {Sandstrom}, {Bigiel}, {Brinks}, {Kramer},
  {Rosolowsky}, {Schuster}, \& {de Blok}}]{Usero2015}
{Usero}, A., {Leroy}, A.~K., {Walter}, F., {et~al.} 2015, \aj, 150, 115

\bibitem[{{Uzuo} {et~al.}(2021){Uzuo}, {Wada}, {Izumi}, {Baba}, {Matsumoto}, \&
  {Kudoh}}]{Uzuo2021}
{Uzuo}, T., {Wada}, K., {Izumi}, T., {et~al.} 2021, \apj, 915, 89

\bibitem[{{Ward} {et~al.}(2022){Ward}, {Harrison}, {Costa}, \&
  {Mainieri}}]{Ward2022}
{Ward}, S.~R., {Harrison}, C.~M., {Costa}, T., \& {Mainieri}, V. 2022, \mnras,
  514, 2936

\bibitem[{{Williams} {et~al.}(2024){Williams}, {Lee}, {Larson}, {Leroy},
  {Sandstrom}, {Schinnerer}, {Thilker}, {Belfiore}, {Egorov}, {Rosolowsky},
  {Sutter}, {DePasquale}, {Pagan}, {Berger}, {Anand}, {Barnes}, {Bigiel},
  {Boquien}, {Cao}, {Chastenet}, {Chevance}, {Chown}, {Dale}, {Deger},
  {Eibensteiner}, {Emsellem}, {Faesi}, {Glover}, {Grasha}, {Hannon}, {Hassani},
  {Henshaw}, {Jim{\'e}nez-Donaire}, {Kim}, {Klessen}, {Koch}, {Li}, {Liu},
  {Meidt}, {M{\'e}ndez-Delgado}, {Murphy}, {Neumann}, {Neumann}, {Neumayer},
  {Oakes}, {Pathak}, {Pety}, {Pinna}, {Querejeta}, {Ramambason}, {Romanelli},
  {Sormani}, {Stuber}, {Sun}, {Teng}, {Usero}, {Watkins}, \&
  {Weinbeck}}]{Williams2024}
{Williams}, T.~G., {Lee}, J.~C., {Larson}, K.~L., {et~al.} 2024, \apjs, 273, 13

\end{thebibliography}

 \appendix
 
\section{CO(3-2) channel maps for NGC~1672}
\begin{figure}
\begin{center}
\includegraphics[clip,width=0.5\textwidth,angle=0]{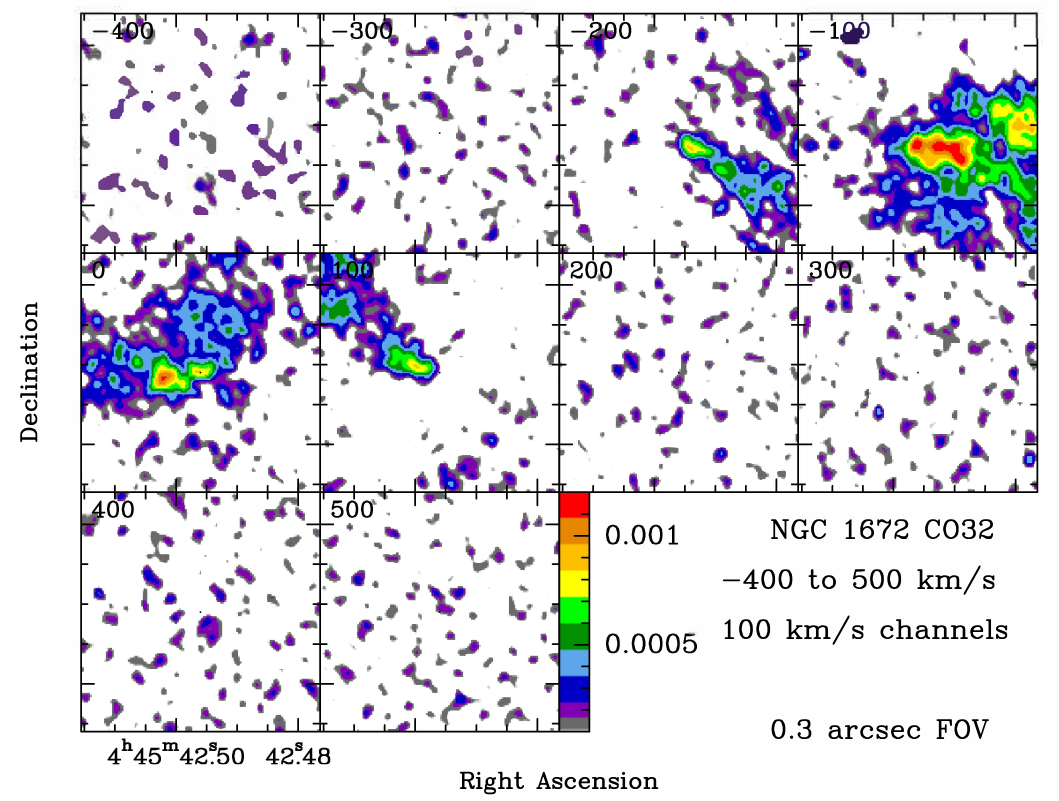}
\vskip+0.0cm
\caption{Channel map in CO(3-2) of NGC~1672.  Each channel is 100~\kms\, wide, and are plotted to the 
the same flux scale, in Jy beam$^{-1}$, averaged over the channel velocity range. Each panel is 
 0.3\arcsec in size.} 
\label{fig:n1672-channels} 
\end{center}
\end{figure} 

 \section{Comparison between low and high resolution velocity maps for HCO$^+$(4-3) }
\begin{figure*}
\begin{center}
\includegraphics[clip,width=0.9\textwidth,angle=0]{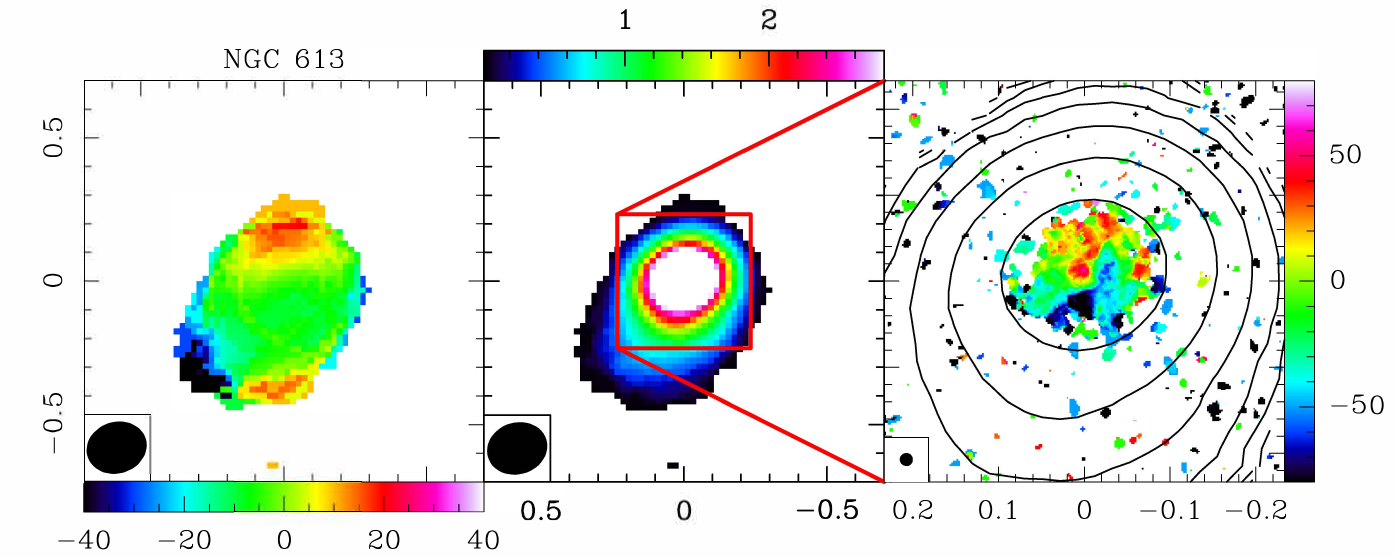}
\vskip+0.0cm
\caption{Low resolution at $\sim$ 0.19\arcsec\, of the velocity field (left), and the density (middle) of the HCO$^+$(4-3) line in NGC~613, compared with the high resolution at $\sim$ 0.015'' (right) of the velocity field. In the right panel, the contours of the low resolution HCO$^+$(4-3) map are superposed.  The color scale is in velocity 
difference from the V$_{sys}$= 1481 \kms.
The spatial scales of all panels are RA-DEC in arc-seconds, from the center of Table \ref{tab:samp}.} 
\label{fig:n613-lo-hires-hco} 
\end{center}
\end{figure*} 

\begin{figure*}
\begin{center}
\includegraphics[clip,width=0.9\textwidth,angle=0]{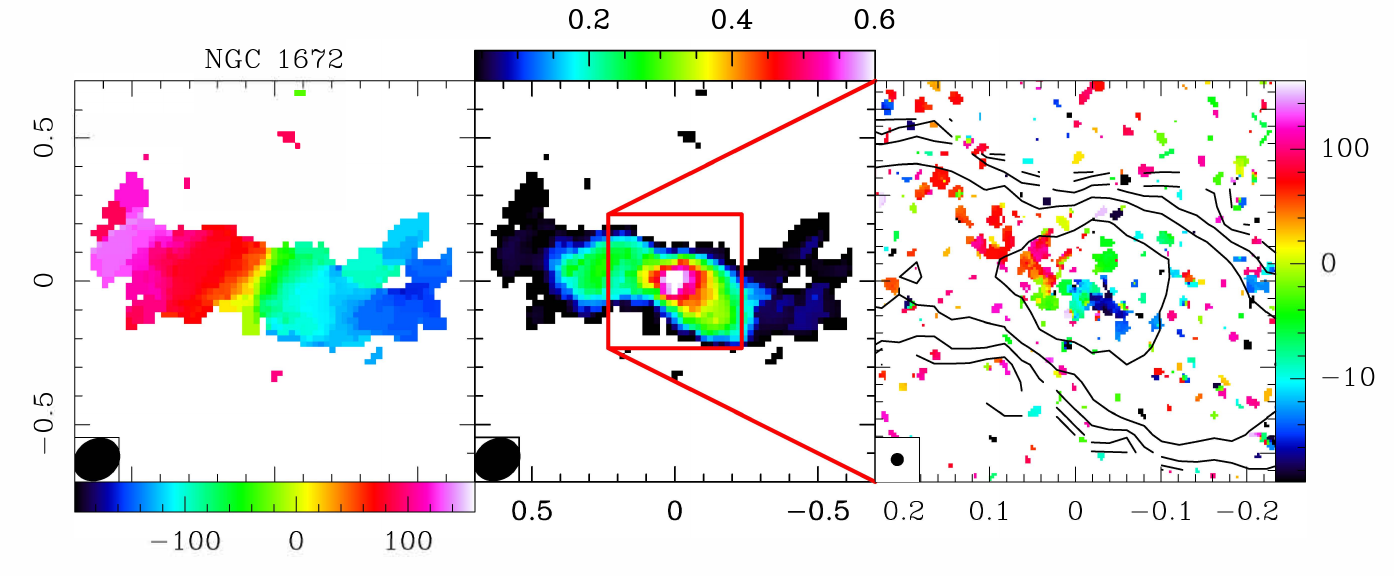}
\vskip+0.0cm
\caption{Same as Fig \ref{fig:n613-lo-hires-hco}, for NGC~1672. The color scale is in velocity difference from the V$_{sys}$= 1331 \kms.} 
\label{fig:n1672-lo-hires-hco} 
\end{center}
\end{figure*}


\end{document}